\documentclass[12pt,a4paper]{article}
\usepackage{pdfsync}
\usepackage{abstract}
\usepackage{amsfonts}
\usepackage{amsmath}
\usepackage{amssymb}
\usepackage{amsthm}
\usepackage{comment}
\usepackage[toc,title,header]{appendix}
\usepackage{enumerate}
\usepackage[top=1.5in, bottom=1.5in, left=1in, right=1in]{geometry} 
\usepackage{graphicx}
\usepackage{tikz}
\usetikzlibrary{arrows,shapes,decorations.text}
\usepackage{epstopdf}
\usepackage[section]{placeins}%disallow floating across sections
\usepackage[pdfstartview=FitB,breaklinks=true,pdfpagemode=UseNone]{hyperref}
\usepackage{mathtools}
\usepackage[comma]{natbib}
\usepackage{picture}
\usepackage{placeins}
\usepackage{setspace}
\usepackage{xcolor}
\usepackage{mathrsfs}
\usepackage{cleveref}
 \usepackage{pxfonts}
 \usepackage[T1]{fontenc}
\usepackage{caption}
\DeclareCaptionLabelFormat{AppendixTables}{B.#2}
\DeclareCaptionLabelFormat{AppendixTable}{C.#2}
\crefname{figure}{figure}{figures}
\creflabelformat{figure}{#2#1#3}
\crefname{equation}{equation}{equations}
\creflabelformat{equation}{#2(#1)#3}
\crefname{lemma}{lemma}{lemmas}
\creflabelformat{lemma}{#2#1#3}
\crefname{theorem}{theorem}{theorems}
\creflabelformat{lemma}{#2#1#3}
\crefname{condition}{condition}{conditions}
\creflabelformat{condition}{#2#1#3}
\crefname{assumption}{assumption}{assumptions}
\creflabelformat{assumption}{#2#1#3}
\crefname{appendix}{appendix}{appendices}
\creflabelformat{appendix}{#2#1#3}
\crefname{enumi}{}{}
\creflabelformat{enumi}{(#2#1#3)}
\crefname{appsec}{Appendix}{Appendices}

\theoremstyle{exampstyle}
\newtheorem{theorem}{\bfseries Theorem }

\newtheorem{assumption}{\bfseries Assumption}

\newtheorem{remark}{\bfseries Remark}

\makeatletter %adjust the space between equation displays and the text
\g@addto@macro\normalsize{%
\setlength\abovedisplayskip{5pt plus 2pt minus 2pt}
\setlength\belowdisplayskip{5pt plus 2pt minus 2pt}
\setlength\abovedisplayshortskip{6pt plus 2pt minus 2pt}
\setlength\belowdisplayshortskip{6pt plus 2pt minus 2pt}
}
\makeatother

\usepackage{titlesec} %adjust the space before and after section header
\titlespacing*{\section}
{0pt}{1ex}{2ex} %first num for indentation; second for above; last for below
\titlespacing*{\subsection}
{0pt}{2ex}{1ex}
\titlespacing*{\subsubsection}
{0pt}{1ex}{1ex}

\usepackage{fancyhdr}
\hypersetup{
	unicode=false,          % non-Latin characters in Acrobat’s bookmarks
	colorlinks=true,       % false: boxed links; true: colored links
	linkcolor=black,          % color of internal links
	citecolor=black,        % color of links to bibliography
	filecolor=magenta,      % color of file links
	urlcolor=black          % color of external links
}
\begin{document}

\title{\large 
%Selection and 
Endogenous Treatment Models with Social Interactions: An Application to the Impact of Exercise on Self-Esteem\footnote{The paper has greatly benefited from comments and suggestions provided by the co-editor James Heckman and two anonymous referees. We also thank Ivan Fernandez-Val, Xiaodong Liu,  \'Aureo de Paula, Alex Poirier, Xun Tang and audiences at Clemson, Georgia Tech and SEA Annual meeting for helpful discussion and comments. We are particularly grateful to John Rust for his detailed suggestions.  All remaining errors are our own.}}
\author{Zhongjian Lin\footnote{John Munro Godfrey, Sr. Department of Economics, University of Georgia. 620 S. Lumpkin St., Athens, GA 30602. Email: zhongjian.lin@uga.edu.} \\
University of Georgia \and Francis Vella\footnote{Department of Economics, Georgetown University. 37th St. N.W. and O Streets N.W., Washington DC 20057. Email: fgv@georgetown.edu.}\\
Georgetown University}
\date{\today}
\maketitle

\begin{abstract}
 We address the estimation of 
%sample selection and 
endogenous treatment models with social interactions in both the treatment and outcome equations.
We model the interactions between individuals in an internally consistent manner via a game theoretic approach based on discrete Bayesian games. This introduces a substantial computational burden in estimation which we address through a sequential version of the nested fixed point algorithm.
%We describe how our methodology is applicable to a large class of models. 
%Our procedure is the first which considers estimation of such a model. 
We also provide some relevant treatment effects, and procedures for their estimation, which capture the impact on both the individual and the total sample. 
Our empirical application examines the impact of an individual's exercise frequency on her level of self-esteem. We find that an individual's exercise frequency is influenced by her expectation of her friends'. We also find that an individual's level of self-esteem is affected by her level of exercise and, at relatively lower levels of self-esteem, by the expectation of her friends' self-esteem. \\

\end{abstract}

\newpage
\begin{doublespace}

\section{Introduction}

%%\textbf{Need to rewrite this paragraph focussing more on the role of peer effects etc} Several methodological papers by James Heckman  \citep[see, for example][]{heckman1974shadow,heckman1978dummy,heckman1979sample} on the
%estimation of sample selection and endogenous treatment models have been greatly
%influential in both the theoretical and empirical microeconometrics literatures. 
%The former
%includes extensions of the original estimators via the relaxation of distributional and parametric assumptions,
%incorporating alternative selection/treatment rules, employing various identification schemes,
%the capacity to estimate models with categorical outcome dependent variables, and the
%adaption of the estimators to a large range of data structures \citep*[see, for
%example,][]{manski1990nonparametric,das2003nonparametric,honore2020selection}. A feature of
%these methodological extensions is their focus on statistical issues rather than those related to economic behavior. 
%This paper addresses an aspect of economic behavior which has been largely ignored but merits attention. Namely, the inherent
%equilibrium and peer effects, capturing the interactions between the individuals whose
%behavior is being modelled, operating in potentially both the selection/treatment and outcome equations. \citep*[see, for example,][]%{manski1993identification,manski2000economic,brock2001discrete,brock2001interactions,heckman2007econom%etric}.\footnote{We use the terms ``peer effects'' and ``social interactions'' interchangeably.}

Despite the substantial growth of empirical investigations focusing on the impact of social interactions in a variety of settings, the role of peer effects has remained relatively unexplored in endogenous treatment and sample selection models. While this partially reflects the absence of suitable data sets, it also represents a shortcoming of this extensive literature as a possible determinant of an individual's behavior is their expectation of their friends'. This suggests that policy makers should explicitly incorporate the role of peer effects in assessing their capacity to encourage individuals to undertake a specific treatment.\footnote{We use the terms ``peer effects'' and ``social interactions'' interchangeably.} Moreover, the effectiveness of a policy is also likely to depend on policy makers' capacity to identify individuals whose behavior most influences the treatment decisions of others. In addition, the outcome targeted by the treatment may also be influenced by peer effects. For example, empirical evidence suggests that there is contagion in both an individual's tendency to exercise and their propensity to be obese. If policy makers were to encourage individuals to exercise to fight obesity, they should incorporate that both the individual's treatment decision and the targeted outcome may both be influenced by their peers' behavior.  As the endogenous treatment model can be interpreted as a special case of the \cite{roy1951some} model, we consider how some estimators for variants of this model can be adapted to account for the presence of peer effects.\footnote{The large literature examining the estimation of variants of the Roy model is motivated by the work of James Heckman \citep[see, for example][]{heckman1974shadow,heckman1978dummy,heckman1979sample}.}

 Endogenous treatment models frequently feature an equation
characterizing the individual's treatment decision and equation(s) describing
the outcome for the respective treated and non treated groups. %\footnote{%
%To simplify terminology we refer to
%the equation describing whether the individual decides to participate or be
%treated as the treatment equation.} 
The presence of endogeneity arises when the unobservables in the treatment equation are
not independent of those in the outcome equation(s) and one cannot 
consistently estimate
the outcome equation without accounting for this. While endogenous treatment models can be sometimes estimated by instrumental variable methods which explicitly purge the treatment variable from the unobservables, they can also be estimated by jointly modeling the treatment and outcome equations while accounting for the correlation between the unobservables across the two equations. This approach requires that  both equations are correctly specified. This is also required if the treatment endogeneity is accounted for by the inclusion of a constructed control function(s) in the outcome equation(s) \citep[see, for example][]{heckman1979sample}.
%Although these models can be  
%estimated jointly by maximum likelihood while incorporating the
%correlation across equations, the more frequently employed approach in
%empirical work is a two-step procedure. The first step estimates
%the treatment equation from which the appropriate control function(s) are
%constructed. The second step estimates the outcome equation(s)
%with the inclusion of the control function as a conditioning variable.

Given this requirement 
%that the treatment equation be correctly specified 
it is notable that the treatment literature generally excludes the presence of peer effects in both the treatment and outcome equations.
%that the individual's treatment
%decision incorporates their expectation of the treatment choices of the members of their social network. 
For example, 
%is the sample selection literature 
in their 
investigation of whether individuals optimally invest in college education
%They
%estimate separate wage equations for
%those attending and not attending college while accounting
%for the selection bias arising from the non college/college choice.
\cite{willis1979education} impose that the
individual's expectation of their network members' college decisions is irrelevant. This
might be problematic if a determinant of an individual's college decision is their
belief about their friends' choices. 
%This will also have implications for estimation if the treatment endogeneity of choice is accounted for by accounting for the correlation between unobservables in the outcome and treatment equations. 
Incorrectly excluding peer effects will result in an inappropriate correction for the endogeneity operating through the dependence in the unobservables.
%However, more importantly, it is critical to understand the role of social interactions in the treatment decision given its implications for public policy purposes.

In addition to assuming that individual's treatment decisions are independent of their peers', it is also assumed that the individual's outcome is not affected by their expectation of their peers' outcomes. While this may be generally reasonable, there are certainly instances when it is not. For example, if the outcome under examination is the individual's wage it is generally sensible to impose that the expectation of their peers' wages is irrelevant. However, if the outcome is a variable with the empirically established feature of contagion, such as obesity or smoking, then it is less reasonable. We examine the estimation of models with peer effects in both the treatment and outcome equations.

%In estimating this type of model it is important to incorporate the endogeneity of the treatment and the role of the peer effects in one or both equation. Estimators motivated in this Roy framework require that the treatment equation is correctly specified and ignoring the peer effects may lead to misspecification. However, more importantly in this endogenous treatment literature it is important for policy purposes that one understands the process generating the treatment decision. This is particularly the case when the treatment decision of oen individual potentially affects the treatment decisions of others in the sample.

In addition to revisiting the estimation of the parameters in the endogenous treatment models in the presence of social interactions, it is also necessary to reconsider the concept of a treatment effect. A treatment effect usually reflects the impact of an individual undergoing the treatment on a specified outcome of interest. However, with social interactions in both treatment and outcomes, the objects related to the impact of undergoing treatment should incorporate how the individuals' treatment decision affects that of the members of their network, both directly and indirectly, and its impact on their outcomes. %Moreover, this also increases the extent of the treatment effect as there is also an effect on the network's outcome from an individual undergoing treatment. 
Another contribution of this paper is to define a treatment effect in this context.

A challenge in introducing "social interactions" into endogenous treatment models is the interdependence between individuals' behavior in both the treatment decision and outcome equations. This requires modeling the interactions between individuals in an internally consistent manner. We employ a game-theoretic approach to capture this simultaneity of the individuals' choices in settings with a large number of individuals in a large social network. The associated computational burden is addressed through two Bayesian games which solve for the simultaneity of peers' choices and outcomes. 
%This is done by projecting
%individual's choice onto a potentially large exogenous space containing the
%relevant individual's characteristics, the individual's friends'
%characteristics, the characteristics of the friends of friends, and so on.
We propose estimation via a sequential version of the fixed point algorithm
\citep[see][]{rust1987optimal,aguirregabiria2007sequential,lin2017estimation}.
%We consider two types of estimators which capture agent interactions in the treatment decision. 
%The first adapts the Heckman (1974) approach and constructs the appropriate likelihood function to jointly model the treatment and outcome equations and the correlation in the unobservables across the two equations. The second corresponds to the \cite{heckman1979sample} two-step estimator which accounts for the inherentendogeneity via the appropriate control function from the treatment equation. 

Our paper makes three contributions. First, we employ a game
theoretic approach to model the individual's treatment choice and outcome in \cite{roy1951some} type models in which the individual's treatment choice and outcome are influenced by
their expectation of their friends' treatment choices and outcomes. While some similar issues are
addressed in \cite*{lee2014binary,lin2017estimation,xu2018social,lin2024binary}, this is the first paper to employ this approach in the endogenous treatment context and to incorporate social interactions in both the treatment and outcome equations. This is a non-trivial and important extension to the current literature. Note that it is straightforward to adjust our proposed procedure if we wish to exclude these interactions  from either equation. Second, we provide treatment effects, and their estimators, which incorporate social interactions in both the treatment and outcome equations. 
%This is interesting as it necessary to provide treatment effects which capture the contagious behavior in both treatment and outcome. 
Finally, we examine how an individual's
level of weekly exercise influences their level of self-esteem when the individual's decision to exercise and their level of self-esteem are potentially influenced by their expectation of their friends' exercise and self-esteem, respectively. 
%This is an important research area given the relationship
%between an individual's self-esteem and their mental, physical and economic
%well being. 
Recent evidence suggests that both exercise and self-esteem have contagious components. Our empirical investigation establishes that exercise has a
positive impact on self-esteem and accounting for its simultaneity increases
this estimated impact. Moreover, we find that an individual's belief of
their peers' exercise frequency has a statistically
significant impact on their own. We also find that an individual's perception of their friends' self-esteem has a direct impact on their own for those with lower levels of self-esteem. %This empirical example highlights the importance of incorporating  peer effects. 
%Our method can also be extended to study triangular relationship in markets, e.g., competitive entry decisions and strategic promotion strategy of firms.

There are three important issues
we do not address as each would greatly complicate the analysis without substantially enhancing the novelty of our approach. First is the nature of the information available
to the agents in forming their beliefs. By assuming the
relevant unobserved information is private, we construct expectations
that are uncorrelated with the other unobservables in the model.
However, in many settings, there may also be publicly shared unobserved information.
This introduces an additional source of endogeneity which we do not account for.
The second is the assumption that the large social networks, and the choice of friends within the network, are 
exogenously determined. While this is a strong assumption it is frequently employed in the social interactions literature \citep*[see, for example,][]{manski1993identification,brock2001discrete,lee2007identification,bramoulle2009identification,calvo2009peer,lee2010specification,goldsmith2013social,lee2014binary}.
The third is our reliance on parametric assumptions. We acknowledge that many of the theoretical advances 
in the endogenous treatment, and the closely related sample selection model, literature are related to the relaxation of parametric assumptions. While many of these advances 
could be applied here, we do not do so as it would greatly complicate estimation. 
%However, we note that our procedure shares some features of the flexibility created by the semi-parametric  approaches.
%Finally, we only allow for
%interactions in the treatment decision and do not incorporate
%them in the outcome equation. 
%While it seems more natural that they appear in the treatment equation, we acknowledge there may be empirical settings in which they should be included in both equations. While we do not formally extend our approach to address these four restrictions we outline below how three of them could be relaxed. However, we delay a detailed investigation of each to future research. The assumption of the exogeneity of the network, in the context of a large network, appears difficult to avoid.

The following section outlines the model and describes the explicit manner in which social interactions
enter the model and discusses the employed Bayesian Nash equilibrium
concept. It also defines the treatment effects associated with our model. Estimation is addressed in \Cref{estimation}. The empirical investigation of the impact of an individual's frequency of physical exercise on their level of self-esteem is described in \Cref{esteem}. \Cref{conclusion} concludes. The proof of our primary result, additional tables and simulation evidence are provided in the Appendices.
\section{A Model of Social Interactions with Treatment Effects}\label{model}
%I THINK WE PROBABLY WANT TO PUT A PEER EFFECT IN THE OUTCOME EQUATION AND THEN SAY WE FIRST CONSIDER THE CASE WHERE ITS COEFFICIENT IS ZERO
%We begin by  describing the model and data structure we consider. We then provide definition of alternative treatment effects in the context of the model. 

\begin{figure}
  \caption{Peer Effects and Treatment Effects }\label{PETE}
\begin{center}
\begin{tikzpicture}[->,>=stealth',shorten >=1pt,auto,node distance=6cm,
                    thick,main node/.style={circle,draw,font=\sffamily\Large\bfseries},roundnode/.style={circle, draw=black, very thick, minimum size=7mm}]
   \node[roundnode, scale=1] at (-4,6) (1) {$D_1$};
       \node[roundnode, scale=1] at (2,6) (2) {$D_2$};

  \node[roundnode, scale=1] at (8,6) (3) {$D_3$};
    \node[roundnode, scale=1] at (-4,8) (4) {$Y_1$};
  \node[roundnode, scale=1] at (2,8) (5) {$Y_2$};
  \node[roundnode, scale=1] at (8,8) (6) {$Y_3$};

  \draw[->](1)--(4);
    \draw[->](2)--(5);
      \draw[->](3)--(6);
        \draw[<->](1)--(2);
          \draw[<->](2)--(3);
            \draw[<->](4)--(5);
              \draw[<->](5)--(6);
 \node[ scale=0.11] at (-6,6) (7) {};
 \node[ scale=0.1] at (10,6) (8) {};
  \node[ scale=0.11] at (-6,8) (9) {};
 \node[ scale=0.1] at (10,8) (10) {};
\draw[->,dashed] (1)--(7);
\draw[->,dashed] (3)--(8);
\draw[->,dashed] (4)--(9);
\draw[->,dashed] (6)--(10);
\def\myshift#1{\raisebox{1ex}}

%\draw [<-,blue,postaction={decorate,decoration={text along path,text align=center,text={|\myshift|Peer Effects}}}] (1) to [bend left] (2);
%\draw [<-,blue,postaction={decorate,decoration={text along path,text align=center,text={|\myshift|Peer Effects}}}] (2) to [bend left] (3);
%
%\draw [->,black,postaction={decorate,decoration={text along path,text align=center,text={|\myshift|Friend Nomination}}}] (1) to [bend right] (2);
%\draw [->,black,postaction={decorate,decoration={text along path,text align=center,text={|\myshift|Friend Nomination}}}] (2) to [bend right] (3);
\end{tikzpicture}
\end{center}
\end{figure}
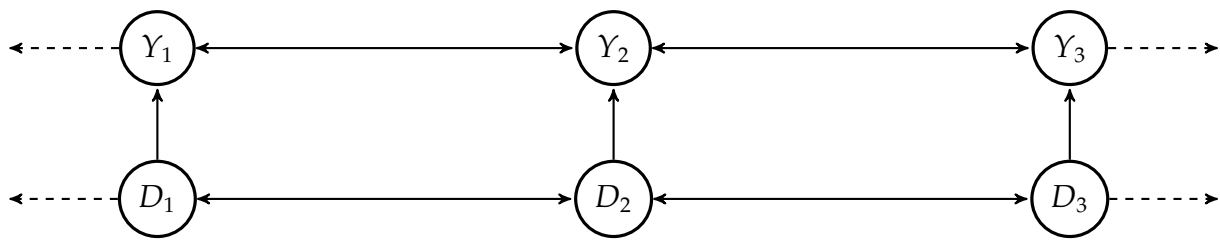

Consider a sample of $i=1,\cdots,n$ individuals for which we observe a binary outcome of interest $Y$, a potentially endogenous binary treatment $D$, and vectors of exogenous variables $X$ and $Z$ which determine the outcome variable and the treatment decisions respectively. For each  $i$ we observe the $j$ members of their network. 
We 
%is assumed to be generated in the presences of peer effects. 
assume the treatment decision and outcome for $i$ are potentially influenced by their expectations of their $j$ peers' treatments and outcomes.

\Cref{PETE} illustrates a simple mechanism governing the peer effects in the treatment choices $D$'s and the outcomes $Y$'s. There are peer effects in the $D$'s as $D_1$ and $D_2$ are interdependent and $D_2$ and $D_3$ are interdependent. For the sake of exposition, we assume  $D_1$ and $D_3$ are not directly related although they are connected through their relationships with $D_2$. These direct dependencies are assumed to operate through social connections or network membership. We allow the same interdependence in outcomes. The dependencies among the $D$'s and $Y$'s are characterized by two Bayesian (discrete) games. 
%Furthermore, we have an endogenous treatment effects from $D$ to $Y$.
\footnote{We restrict attention to the own treatment effects. Previous work has considered alternative exposure to characterize the dependence, e.g., own $D$ and average friends' $D$ affecting $Y$, \citep[see][]{manski2013identification,aronow2017estimating,leung2022unconfoundedness}.}

In addition to the direct treatment effect from $D_i$ on $Y_i$, the peer effects in $D$ and $Y$ introduce additional general equilibrium effects. The dependence across $D$'s means that a shock to individual $i$ could not only affect $i$'s treatment decision, but also their friends' treatment decisions, the friends of their friends' treatment decisions, and so on. Similarly, if there is a direct treatment effect from $D_i$ on $Y_i$, it is possible that the change in $Y_i$ will affect their friends' outcomes, the friends' of their friends' outcomes, and so on. %In defining some measures of treatment effects, it is necessary to incorporate this consideration.
%rather than taking simple average potential outcome changes across all individuals. Furthermore, the peer effects in outcome make the calculation of the treatment effects even more complicated that the impact from $D_i$ on $Y_i$ is not a simple closed form formula, but the final value of $Y_i$ comes from the equilibrium, i.e., value of fixed points.
Below we define treatment effects in the context of a social network.

\subsection{Econometric Model}
%\textcolor{blue}{It is not obvious $\mu_Y$ and $\mu_D$ deliver the statement that we can use different cutoffs to define $Y$. In our empirics, we fixed the definition of $D$ with less or equal to 5 times a week as 0, and more than 5 as 1. There is no need to have $\mu_D$.}

The econometric model has the form:
%\textbf{We need to explain what types of D and Y we can handle.....Obviously we can handle ordinal treatments and outcomes but more complicated.....Presumably we can handle non order multinomial but even more complicated. However, we cannot handle continuous treatments or outcomes with peer effects....we need to explain why...Do we need to explain idetification or do we just identification comes from inclusion of peer effects in treatment equation}
\begin{align}
\label{outcome}Y_{i} &=1\{X_{i}'\beta _{O}+\gamma D_{i}+\delta W_i^*+u_{i}>0\} \\
\label{selection}D_{i} &=1\{Z_{i}'\beta ^T+\alpha V_{i}^*+v_{i}>0\} 
\end{align}%
where %$X$ and $Z$ are vectors ofexogenous variables; 
the $\beta ,$ $\gamma $, $\delta$ and $\alpha $ are parameters; $%
u$ and $v$ are potentially correlated error terms and $W_i^*$ and $V_i^*$ are constructed variables capturing the individual's beliefs of their friends' values of $Y$ and $D$. These two terms incorporate social interactions and are determined within the model. As their values directly depend on the model's parameters, the
%The $q$ and $k$ subscripts denote that different parameters are estimated at different quantiles and the $\mu$'s denote thresholds of either latent or observed dependent variables.
%We assume each individual $i$ has $N_{i}$ direct friends which also appear in the sample. 
%The $W_i^*$ and $V_i^*$ are assumed to be 
%described below and they 
%known functions
%of the expectation of the $Y_{j}'$s and $D_{j}^{\prime }s$ where the individuals denoted
%with the subscript $j$ are members of $i^{\prime }s$ network. These expectations are functions of the parameters of the model and 
$ ^*$ denote that they are evaluated at the equilibrium values. As these social interactions variables can also be evaluated at non equilibrium parameters values, we drop the $^*$ when doing so.
We allow the exogenous explanatory variables to differ across the two equations but note that $X$ and $Z$ may be identical as the model is identified through parametric assumptions and the inclusion of the peer effects in the treatment equation.

We construct the $W_i^*$ and $V_i^*$ via a game theoretic
approach. Let $F_{ij}=1$ denote that individual $i$ considers $j$ a friend.
We set $F_{ii}=0$ by convention and denote $F_{i}=\{j\in I:F_{ij}=1\}$ as
the set of $i^{\prime }s$ peers. The number of $i^{\prime }s$ friends is
$N_{i}=\#(F_{i})$. We assume $u_{i}$ and $v_{i}$
represent private information only known to agent $i$. For the outcome equation, $\mathbb I_O\equiv\{X_i,Z_i,D_i,F_i\}_{i=1}^n$ denotes the public information, while for the treatment equation, it is denoted by  $\mathbb I_T\equiv\{Z_i,F_i\}_{i=1}^n$.%The remaining
%sources of information (denoted as $\mathbb{I}=\{X_{i},Z_{i},F_{i}\}_{i\in
%I} $) are treated as public. 
 The private information
assumption is important not only for the construction of beliefs but
also because of its implications for the endogeneity of the treatment
decision. Even with treatment endogeneity the $W_i^*$ and $V_i^*$ can be treated as exogenous.

We use a Bayesian game to characterize the 
social interactions in each of the equations. We follow \cite{manski2000economic} to conceptualize individuals as decision makers endowed with preferences, who form expectations and face constraints. Individuals make simultaneous treatment choices. With incomplete information,
individuals do not observe the actions of their peers but form a belief of
them \citep*[see, for example,][]{brock2001discrete,lee2014binary,lin2017estimation,xu2018social,lin2024binary}. The belief, $E(D_{j}|\mathbb{I}_T)$'s, rather than the observed treatment choices, $D_{j}$'s,
affect $i^{\prime }s$ decision. Similarly, $E(Y_{j}|\mathbb{I}_O)$'s affects the individual's outcome. The incomplete information structure facilitates a tractable solution to network games with a large number of players. It also accommodates the imposition of the primitive conditions required to establish the contraction mapping property and uniqueness of the equilibrium. We elaborate on these conditions below.

We adopt a first-order Markovian setting in which only direct friends deliver peer effects and others have an indirect impact through friendship links. This is reflected in \Cref{PETE}. The first-order sequence seems reasonable noting that the use of higher-order Markovian sequences is relatively undeveloped in this literature. 

We construct the social interactions terms as: 
\begin{align}
  W_{i}^*&=\frac{1}{N_{i}}\sum_{j\in F_{i}}E(Y_{j}|\mathbb{I}_O),\\
  V_{i}^*&=\frac{1}{N_{i}}\sum_{j\in F_{i}}E(D_{j}|\mathbb{I}_T).  
\end{align} This represents a form of the summary statistic regarding the belief of 
$i^{\prime }s$ peers' behavior, which ignores their identities.\footnote{We adopt a local average setting and impose homogenous peer effects. We could allow alternative characterizations of $W$ and $V$, such as heterogenous peer effects, e.g., in \cite{lin2017estimation}, where the magnitudes of the peer effects are partially determined by relative centrality.} This approach allows a tractable equilibrium characterization of
the simultaneous treatment and outcome choices, while the Bayesian Nash equilibria (BNE)
%(contraction fixed point)
facilitate estimation in the presence of a high dimensional exogenous space. This characterization of peer effects is relatively simple as it captures the average expected behavior of the members of the individual's network. One could also employ alternative measures but this representation seems reasonable.

The use of $E(D_{j}|\mathbb{I}_T)$'s and $E(Y_{j}|\mathbb{I}_O)$'s, rather than the observed $%
D_{j}$'s and $%
Y_{j}$'s, reflects that individuals are
simultaneously making decisions and do not see the choices of the others in
the network. This private information assumption implies the endogeneity
of the peers' choices operates through these expectations. This excludes unobserved factors which influence both the individual's and
their peers' treatment choices and outcomes. Including the observed $Y$ and $D$ imposes that the individual's treatment choices and outcomes are determined by the actual treatment choices and outcomes of her friends. This might suggest there are unobserved factors which influence both the
choices of $i$ and the other network members. This would
result in the choices being endogenous even after the inclusion of $W_{i}^*$ and  $V_{i}^*$.
%We do not formally address this issue here although we outline below a way for incorporating shared unobserved information in the private information setting (DID WE KKEP THIS?). This has the cost of greatly complicating the analysis and so 
We highlight that this private information assumption is commonly adopted in analyses of large network games. See, for example, 
\cite*{lee2014binary,lin2017estimation,xu2018social,lin2024binary}. There is also a large number of papers in
theoretical microeconomics which study 
incomplete information games \citep[see, for
example,][]{morris2003global,bergemann2013robust} and the theoretical econometrics literature on discrete games frequently employs an
incomplete information structure \citep*[see, for example,][]
{seim2006empirical,aguirregabiria2007sequential,pakes2007simple,aradillas2010semiparametric,doraszelski2010computable,aradillas2012pairwise,aradillas2020econometrics,de2012inference,wan2014semiparametric,dai2015regulation,lewbel2015identification,lin2021multiplex}. 

The Bayesian game with incomplete information structure provides a tractable characterization of the large network game. Moreover, as its equilibrium is characterized in terms of conditional choice probabilities (CCPs), it facilitates the imposition of primitive conditions required for the existence of a contraction mapping and a unique equilibrium. This is not possible in the complete information setting. As the number of players is growing in the large network game, it is infeasible to investigate the case of multiple equilibria. Tractability is attractive for the structural analysis of network games noting they have been successfully employed to study
strategic interactions among agents in a variety of economic and social
situations.

Define $P_T^*=(P_{T1}^*,\cdots,P_{Tn}^*)\equiv(P(D_1=1|\mathbb I_T),\cdots,P(D_n=1|\mathbb I_T))$ and $P_O^*=(P_{O1}^*,\cdots,P_{On}^*)=(P(Y_1=1|\mathbb I_O),\cdots,P(Y_n=1|\mathbb I_O))$ recalling that $^*$ denotes the equilibrium values. We make the following assumptions to establish the uniqueness of the two Bayesian
games.
\begin{assumption}\label{normalization}
The error terms $u_i$ and $v_i$ are independent across individuals and are bivariate normally distributed with variances $\sigma^2
_{u}=\sigma^2 _{v}=1$ and correlation $\rho$. The $(u_{i},v_{i})$ are independent of $(X_{i},Z_{i})$. 
\end{assumption}

Under Assumption 1 the CCPs of the treatment decisions are given as:
\begin{equation}\label{bnes}
  P_{Ti}^*=  P(D_i=1|\mathbb I_T)=P(Z_i'\beta_T+\alpha V_i^*>-v_i)=\Phi\Big(Z_i'\beta_T+\alpha V_i^*\Big), i=1,\cdots,n,
\end{equation}
where $\Phi$ denotes the standard normal cumulative distribution function. As the equilibrium value of $V_i$, $V_i^*$, is defined on  $P(D_j=1|\mathbb I_T), j\in F_i$, \Cref{bnes} characterizes the Bayesian Nash equilibrium, in terms of CCPs, in a nonlinear simultaneous system of $n$ equations.

For the outcome decision, since $D_i\in \mathbb I_O$, we have: 
\begin{align}
\begin{split}\label{bneo}
    P(Y_i=1|\mathbb I_O)&=D_iP(Y_i=1|\mathbb I_O\backslash D_i,D_i=1)+(1-D_i)P(Y_i=1|\mathbb I_O\backslash D_i,D_i=0)\\
    &=\frac{D_iP(Y_i=1,D_i=1|\mathbb I_O\backslash D_i)}{P(D_i=1|\mathbb I_T)}+\frac{(1-D_i)P(Y_i=1,D_i=0|\mathbb I_O\backslash D_i)}{P(D_i=0|\mathbb I_T)}\\
    &=\frac{D_i\Phi_2(X_i'\beta_O+\gamma D_i+\delta W_i^*,Z_i'\beta_T+\alpha V_i^*;\rho)}{\Phi\Big(Z_i'\beta_T+\alpha V_i^*\Big)}+\frac{(1-D_i)\Phi(X_i'\beta_O+\gamma D_i+\delta W_i^*)}{1-\Phi\Big(Z_i'\beta_T+\alpha V_i\Big)}\\
    &~~~~-\frac{(1-D_i)\Phi_2(X_i'\beta_O+\gamma D_i+\delta W_i^*,Z_i'\beta_T+\alpha V_i^*;\rho)}{1-\Phi\Big(Z_i'\beta_T+\alpha V_i^*\Big)}, i=1,\cdots,n,
\end{split}
\end{align}
where $\Phi_2$ denotes the standard bivariate normal cumulative distribution function and $W_i^*$ represent the equilibrium value of $W_i$.

We prove the uniqueness of these solutions below. To do so, we denote $V_i(P_T)\equiv \frac{1}{N_i}\sum_{j\in F_i}P_{Tj}$ and $W_i(P_O)\equiv \frac{1}{N_i}\sum_{j\in F_i}P_{Oj}$ and 
define the best response functions:
\begin{equation}\label{brfs}
\Psi_{Ti}(\theta;P_T)\equiv\Phi(Z_i'\beta_T+\alpha V_i(P_T)),i=1,\cdots,n,
\end{equation} and 
\begin{align}\label{brfo}
\begin{split}
  \Psi_{Oi}(\theta;P_O)\equiv&\frac{D_i\Phi_2(X_i'\beta_O+\gamma D_i+\delta W_i(P_O),Z_i'\beta_T+\alpha V_i;\rho)}{\Phi\Big(Z_i'\beta_T+\alpha V_i\Big)}\\
    &+\frac{(1-D_i)\Big[\Phi(X_i'\beta_O+\gamma D_i+\delta W_i(P_O))-\Phi_2(X_i'\beta_O+\gamma D_i+\delta W_i(P_O),Z_i'\beta_T+\alpha V_i;\rho)\Big]}{1-\Phi\Big(Z_i'\beta_T+\alpha V_i\Big)},
\end{split}
\end{align}
and $\psi_{Oi}\equiv \frac{\partial \Psi_{Oi}(\theta;P_O)}{\partial W_i}$.

\begin{assumption}\label{msi}
(i) There exists $M > 0$ such that $0 < \max N_i<M$. (ii) The social
influences are moderate; i.e., $0\leq\alpha<\sqrt{2\pi}$ and $|\delta|<\frac{1}{\sup\psi_{Oi}(\cdot)}$.
\end{assumption}
\begin{remark}
    \Cref{msi} (i) limits the number of friend nominations and generates a sparse network when the network size goes to infinity. This creates subnetworks that are nearly independent. This weak dependence is important for inference in networks. \Cref{msi}(ii) is required to  show the mappings are contractive in the two Bayesian games of $D$'s and $Y$'s. It conveys that there are dependencies among friends' treatment choices and outcomes, but that friends' choices and outcomes are not dominant.
\end{remark}

\begin{theorem}\label{unique}Suppose \Cref{normalization,msi} holds, there exists a unique Bayesian Nash equilibrium in both games represented by \Cref{bnes} and \Cref{bneo}.\footnote{The proof of this theorem is provided in Appendix A.}
\end{theorem}

The multiplicity of the game hampers the coherency of econometric models \citep[see][]{tamer2003incomplete}. When the number of players is small, the total
number of equilibria is also small and a partial identification approach
enables interval estimation  \citep*[see, for example,][]
{tamer2003incomplete,chernozhukov2007estimation,ciliberto2009market,tamer2010partial,balat2020multiple,ciliberto2020market}. When the number of players is an increasing $n$, the number of equilibria
goes exponentially to infinity, and partial identification is not sufficient
or possible for inference. In a large network, data from one equilibrium is observed even though multiple equilibria exist. 

As our model features a binary treatment and a binary outcome it may initially appear somewhat restricted. However, our modeling approach can be extended to both ordered ordinal treatments and outcomes with the added complication of defining the appropriate network effects. It could also be extended to unordered multinomial treatments and outcomes with the associated increase in computation. The model, however, cannot have a continuous outcome measure if there are peer effects in the outcome equation.  \cite{dai2015regulation} propose a discrete game for continuous outcomes in a competition framework. They consider repeated independent markets, represented by games with a small number of players, and estimate the model via a two-step semiparametric estimation method. In our large network game, the conditional expectation of the continuous outcomes has the public information set as conditional covariates whose dimension is growing with the sample size. This prohibits nonparametric estimation of the CCPs. Furthermore, the characterization of the equilibrium is challenging and we are unable to provide a tractable format that facilitates estimation.

We can handle a form of the continuous treatment and outcome case by transforming these continuous variables into binary variables. For example, consider where the treatment and outcome variables are both continuous. We could estimate the model's parameters via our methodology after creating binary outcomes on the basis of whether the observed variables were above a specific threshold such as, for example, some specific quantile. The estimated parameters would not have the same interpretation as the original model but are nevertheless interesting. As our empirical example features continuous measures of treatment and outcome we illustrate this below. This allows the parameters to change depending on the values of the thresholds in the definitions.
%and is similar to distribution regression. 

\subsection{Treatment Effects}
%\textbf{I think we need to extend this section and make it more interesting.....are there any other type of treatment effects we can consider here......For example, is there a way we change the behvaior of an individual in such a way that it changes the behavior of others? presumably this has to operate thru the treatment choice probabilities but we want a way of affecting the behavior of others by changing the behavior of one or more persons}

%\textbf{We also need to explain that even without the above complication we can estimate a series of treatment effects corresponding to different defintions of Y and D..We look at both above median but we can define as many as we want }
A number of the model's parameters are economically interesting. The $\delta$ and $\alpha$  capture the peer effects and are relatively straightforward to interpret. %For example, if $\mu_{Y}$ and $\mu_{D}$ denote the mean of $Y$ and $D$ respectively, $\alpha$ describes how the expectation of the average treatment decision of the individual's peers affects their own treatment decision. 
The $\delta$ parameter describes how the expectation of the average outcome of the individual's peers' outcome affects their own outcome. We highlight again that while the estimation of $\alpha$ follows from earlier work we are the first to provide a procedure for $\delta$ in the presence of an endogenous treatment decision with peer effects. The impact of changes in treatment status $D_i$'s are not isolated and must incorporate the multiplier effects operating through the equilibrium conditions. The treatment effects thus depend on the magnitudes of $\gamma$, $\alpha$ and $\delta$. 

%For economic policy the more interesting implications operate through the $\gamma$ parameter. 
%We consider two types of treatment effects.The first, which we refer to as a partial treatment effect, is related to the change in the treatment decision resulting from fluctuations in the private information $v$. This is not associated with any general equilibrium effects as the shock to the model reflects information only available to individual $i$. While this is a relatively limited object given the richness of the model, it corresponds to the conventional understanding of a treatment effect. The second is the more interesting treatment effect and captures a policy shock to $i$ which also affects the behavior and outcome of the $j$ friends in $i$'s network. This type of policy shock affects public information and induces general equilibrium effects. 

We consider two types of treatment effects.
The first, which we refer to as a partial treatment effect, is related to the change in the treatment decision of individual $i$ while holding the treatment decisions of other individuals constant. For example, this corresponds to a change in the treatment status resulting from a change in the value of the private information $v_i$. This is not associated with any general equilibrium effects in the treatment equation as the change in status is information only available to individual $i$. However, it does allow for changes in the outcome decisions of other individuals if their outcomes respond to a change in  their expectation of $i$'s outcome.  
The second is the more interesting treatment effect and corresponds to a change in the treatment status of $i$ resulting from a change in the index value in the CCP of the treatment equation. This could reflect a change in the values of the conditioning variables or the parameters in the index. In addition to potentially changing the treatment choice and outcome of individual $i$ it may also affect the treatment choices and outcomes of the $j$ friends in $i$'s network. This change in treatment status affects public information and induces general equilibrium effects. 

%\textcolor{red}{Im not sure this pargraph is clear enough}
The partial treatment effect 
%ignoring the peer effects, 
%operates through $\gamma$ and 
corresponds to $P(Y_i=1|\mathbb I_O,D=1)-P(Y_i=1|\mathbb I_O,D=0)$. 
%In the presence of peer effects, the change of $D$ from $0$ to $1$ potentially affects the value of equilibrium term $W$. 
We calculate it by employing the estimated parameters to evaluate the CCPs at $D_i=0$ and $D_i=1$ for a specific individual while holding the $D$'s for the other individuals constant. Denote $D_{-i}\equiv\{D_j:j\neq i\}$. %\textcolor{blue}{Can we define a treatment effect from changing the index which captures i) the network treatment effect ii) the total effect operating through all the network effects?}
We consider the impact of $D$ going from $0$ to $1$ from a change in $v$ for individual $i$. 
%This type of change would not affect the CCPs in the Bayesian game of $D$'s. 
More explicitly, denote the equilibrium CCPs profile associated with $(D_i=0, D_{-i})$ as $P^0$ and that associated with $(D_i=1,D_{-i})$ as $P^1$.\footnote{The equilibrium profiles would be different for different individuals. For simplicity, we use $P^1$ and $P^0$ for the two equilibria profiles for all individuals.} The 
%(marginal) 
partial treatment effect for individual $i$ is calculated as: 
%\textcolor{blue}{do we need to explain how you obtain this?} as: It is straightforward calculation.
\begin{align}\label{me}
\begin{split}
PTE_i\equiv&P(Y_i=1|D_i=1,\mathbb I_O)-P(Y_i=1|D_i=0,\mathbb I_O)=\frac{\Phi_2(X_i'\beta_O+\gamma+\delta W_i(P^1),Z_i'\beta_T+\alpha V_i^*;\rho)}{\Phi\Big(Z_i'\beta_T+\alpha V_i^*\Big)}\\
&-\frac{\Phi(X_i'\beta_O+\delta W_i(P^0))-\Phi_2(X_i'\beta_O+\delta W_i(P^0),Z_i'\beta_T+\alpha V_i^*;\rho)}{1-\Phi\Big(Z_i'\beta_T+\alpha V_i^*\Big)}.
\end{split}
\end{align}
Averaging $PTE_i$ over all individuals provides an estimate of  the average partial treatment effect:
\begin{equation}\label{ate}
APTE=\frac{1}{n}\sum_{i=1}^nPTE_i.
\end{equation}

As noted above, the $APTE$ holds the treatment status of all individuals other than $i$ constant. However as the $APTE$ does incorporate the change in $i$'s outcome which results from the other individuals' outcome responses to $i$'s outcome response, and individual $i$'s response to those changes, it captures the equilibrium effects in the outcome equation only. In evaluating the $ATPE$ we only include the effect on $i$'s outcome resulting from a change in their treatment status noting that this does include the effect on $i$'s outcome resulting from the interaction between individuals in the outcome game.

%\textcolor{blue}{Now consider a change in the treatment status of $i$ which operates through a change in the CCP treatment index. This will have general equilibirum effects as it has capacity to change the treatment status of other individals, excluded in the $ATPE$, with its implications for changes in the outcome choices and the two equilbiurm effects operating though the two Baysian games. } 

Now consider the impact of a change in $D_i$ resulting from variation in the index value inside the CCP.  
%resulting from a change in the value of one the variables in the index. 
This is likely to result in a different equilibrium in the Bayesian game of $D$'s as the  $P_T^*$ will vary with different index values. Denote the initial CCPs in the treatment equation as $P_T^0$ and those associated with the new index value as $P_T^1$. 
%That is, there is a general equilibrium treatment choices. 
To estimate the impact of this movement in the index value we must calculate the composite individual treatment effects which comprises of the changes in the individuals's outcome, the outcomes of their friends, the outcomes of the friends of their friends, etc., resulting from the new index value in the CCP of $i$ which produces variation in the treatment status of any other individual. This follows as the CCPs in the Bayesian game in \Cref{brfo} also depend on the values of $D_{-i}$, and the $D_i$'s depend on the values of the CCPs, the covariates and the unobserved private error shock. Given the nature of this composite effect, it is necessary to evaluate the changes of the $D_{-i}$ of all individuals other than $i$ in the network. This can be done by simulating the random errors $v_{-i}$'s in the treatment choice equation. Denote the CCPs for the $Y_i$'s associated with $D_i=1$ for the $b$-th simulation as $P_O^{1b}$ and $P_O^{0b}$. Each simulation provides an estimate of the individual treatment effects. We then define the composite treatment effects as the average over the $B$ simulations as:%\ \textcolor{blue}{it is not clear what the B simulations is capturing } 

\begin{align}\label{cte}
\begin{split}
CTE_i\equiv&\frac{1}{B}\sum_{b=1}^B\frac{\Phi_2(X_i'\beta_O+\gamma+\delta W_i(P_O^{1b}),Z_i'\beta_T+\alpha V_i(P_T^1);\rho)}{\Phi\Big(Z_i'\beta_T+\alpha V_i(P_T^1)\Big)}\\
&-\frac{\Phi(X_i'\beta_O+\delta W_i(P_O^{0b}))-\Phi_2(X_i'\beta_O+\delta W_i(P_O^{0b}),Z_i'\beta_T+\alpha V_i(P_T^0);\rho)}{1-\Phi\Big(Z_i'\beta_T+\alpha V_i(P_T^0)\Big)}.
\end{split}
\end{align}

This highlights that a change in $D_i$ from 0 to 1 resulting from variation in the value of the CCPs in the treatment participation game has three effects: 1) Direct Effects captured by $\gamma$; 2) Indirect effects resulting from the changes in the peer effects in the treatment equation operating through $V_i^*$ due to the new values in  CCPs in the treatment equation, i.e., from $P_T^0$ to $P_T^1$; 3) Indirect effects resulting from the different values of the peer effects in the outcome equation through $W_i^*$  due to the updated CCPs in the equilibrium of the outcome equation.

   % Consider a change in $D$ from 0 to 1 resulting from a change in the value of the CCPs in the treatment participation game. There are three channels through which the treatment effects will now operate: 1)Direct Effects captured by $\gamma$; 2) Indirect effects resulting from the changes of the peer effects in the treatment equation operating through $V_i$ due to the changes of CCPs in treatment equation, i.e., change from $P_T^0$ to $P_T^1$; 3) The changes of CCPs in the outcome game i.e., change from $P_O^0$ to $P_O^1$.

The treatment effects above only calculate the impact on the focal individual.  As there are possible changes of the treatment status of all individuals this may result in double counting. While the $APTE$ can still be computed while ignoring the outcome responses of other individuals, it is less reasonable to do so for the composite effects. In \Cref{counterfactual} we compute composite treatment effects resulting from some index changes in the treatment decision equation under different scenarios and illustrate aggregate outcomes. We interpret these different scenarios as counterfactuals as they correspond to situations in which we influence the treatment decisions of a targeted group of individuals \citep[see, for example,][]{galeotti2020targeting,sun2023structural}.  

%\textcolor{blue} {we need to have a longer discussion here of what this means...that is the treatment is calculated for the whole economy for changing individual i's CCP...presumably the individual treatment effect is bounded between 0 and 1, the network effect between 0 and J, and the total effect between 0 and N......how should we best measure these?}

 %A change in individual $i$'s treatment decision has the potential to influence the behavior of every individual in the sample depending on the "strength" of the effect and the structure of the network. 
 %For example, the total treatment effect resulting from a change in $i$'s treatment decision will have the direct effect on $i$'s which we call the individual's direct treatment effect, the effect on the members of $i$'s network which we will call the network treatment effect, and the subsequent effect on the remaining members of the sample operating through the networks of the networks of the networks...etc. 
 %In the empirical section, we examine how they differ across different individuals in the sample.
\subsection{Selection Models}
While our primary focus is on estimating treatment effects, our procedure is also applicable to the sample selection literature motivated by Heckman (1974,1979). However, for these models, it is necessary to restrict attention to the special case of no peer effects in the outcome equation. This is because going from the full sample to the selected sample will result in the potential loss of the selected sample's peers. Thus while we are able to create the $V_i^*$, there are potential difficulties in the construction of $W_i^*$. Accordingly, we focus on the case where there are no peer effects in the outcome equation. This restriction is reasonable in many empirical applications. Moreover, the exclusion of $W_i^*$ from the outcome equation has the advantage that we can handle models with continuous outcomes.  The Willis and Rosen (1979) study noted above is an example where these two features are applicable. That is, the selected outcomes are continuous, namely wages, and it is reasonable to assume there is no role of peer effects in wages.

%An industrial organization example in which our method can be employed is the study the impact of competition in markets. For example, to examine market entry and the subsequent promotion strategy. In this setting  we could have the entry decision as $D$ and whether to intensively promote after entry as $Y$. For both the entry and promotion decisions, there are strategic interactions among firms/brands in the market.

\section{Estimation}\label{estimation}
\subsection{Model Parameters}\label{Model Parameters}
We now propose a sequential method, which we refer to as Nested Pseudo Joint Likelihood (NPJL),  to estimate the model parameters. Since the equilibria are based on a game with $n$ players, we illustrate the equilibrium characterization and the asymptotic analysis by using a subscript in the choice probabilities profile to denote their dimension, i.e., $P_{[n]}$. The CCPs of $D_i$'s and $Y_i$'s are then denoted as $P_{Ti[n]}$ and $P_{Oi[n]}$. Denote $P_{[n]}=(P_{T[n]}',P_{O[n]}')'=(P_{T1[n]},\cdots,P_{Tn[n]},P_{O1[n]},\cdots,P_{On[n]})'$. Accordingly, we denote $\Psi(\cdot)=(\Psi_{T1},\cdots,\Psi_{Tn},\Psi_{O1},\cdots,\Psi_{On})'$. When the CCPs in \Cref{bnes,bneo} are known, we can estimate the recursive probit model by maximum likelihood. However, when the CCPs are solved in fixed points, the high dimension of the conditioning set does not accommodate CCP estimation \citep{hotz1993conditional}. We
then define the following pseudo likelihood function:
\begin{equation}\label{pll}
   L_n(\theta;P_{[n]})= L_n(\theta;P_{[n]})=\frac{1}{n}\sum_{i=1}^nl_i(\theta;P_{[n]}),
\end{equation}where $l_i(\theta;P_{[n]})$ is the pseudo log conditional probability of $(Y_i=y_i,D_i=d_i)$, e.g., $l_i(\theta;P_{[n]})=\log \Phi_2(X_i'\beta_O+\gamma D_i+\delta W_i(P_{O[n]}),Z_i'\beta_T+\alpha V_i(P_{T[n]});\rho)$ when $y_i=d_i=1$. 

The model can now be estimated in the following iterative manner:

\begin{enumerate}
\item \textbf{Initiation:} Conjecture the CCPs
in the Bayesian Nash equilibria for both the treatment and outcome equations. For example, estimate the model by probit while assuming exogenous treatment and peer effects. Denote these as $P^{(0)}=(P_O^{(0)},P_T^{(0)})$. Generate $W_i(P_O^{(0)})=\frac{1}{N_i}\sum_{j\in F_i}P_{Oj}^{(0)}$ and $V_i(P_T^{(0)})\equiv \frac{1}{N_i}\sum_{j\in F_i}P_{Tj}^{(0)}$.

\item \textbf{Iteration:} Given $W_i(P_O^{(K)})$ and $V_i(P_T^{(K)})$, estimate $\hat{%
\theta}^{(K+1)}$ via bivariate probit using the likelihood $L(\theta;P^{(K)})$. Update the choice
probabilities in each iteration as: 
\begin{align*}
P^{(K+1)}=\Psi\Big(\hat{\theta}^{(K+1)};P^{(K)}\Big).
\end{align*}

\item \textbf{Convergence:} Iterate until the specified convergent criterion
based on $\Vert P^{(K)}-P^{(K+1)}\Vert $ is satisfied. Denote the K-th
estimate as $\hat{\theta}$ and $\widehat{P}_{[n]}$.
\end{enumerate}

To derive the asymptotic properties of this estimator it is necessary to incorporate the dimensions of its various components. Denote $\theta _{0}$ as the true parameter. Let: 
\begin{align*}
L_{0}(\theta;P_{[n]})& \equiv \mathbb{E}\big[l_i(\theta ;P_{[n]})\big], \\
\tilde{\theta}_{0}(P_{[n]})& \equiv \underset{\theta \in \Theta }{\arg \max }~~%
L_{0}(\theta;P_{[n]}), ~~\psi _{0}(P_{[n]}) \equiv \Psi (\tilde{\theta}_{0}(P_{[n]});P_{[n]}), \\
\tilde{\theta}_{n}(P_{[n]})& \equiv \underset{\theta \in \Theta }{\arg \max }~~%
L_n(\theta;P_{[n]}), ~~\psi _{n}(P_{[n]}) \equiv \Psi (\tilde{\theta}_{n}(P_{[n]});P_{[n]}).
\end{align*}%
Define the population NPJL fixed points set as $\Lambda _{0n}\equiv \{(\theta
,P_{[n]})\in (\Theta ,\mathcal{P}_n):\theta =\tilde{\theta}(P_{[n]}),P_{[n]}=\psi _{0}(P_{[n]})\}$ and
the NPJL fixed points set of sample size $n$ as $\Lambda _{n}\equiv
\{(\theta ,P_{[n]})\in (\Theta ,\mathcal{P}_{n}):\theta =\tilde{\theta}_{n}(P_{[n]}),P_{[n]}=\psi
_{n}(P_{[n]})\}$. Let $\mathcal{N}$ denote a closed neighborhood of $(\theta
_{0},P^{\ast }_{[n]})$. Let $\widetilde{Z}_{i}\equiv (Z_{i}^{\prime },V_{i}^*)$ and $\widetilde{X}_{i}\equiv (X_{i}^{\prime },D_i,W_{i}^*)$.

To establish the $%
\sqrt{n}$ consistency and asymptotic normality of $\hat{\theta}$, we impose the following rank and regularity conditions.
%\textcolor{blue}{dont we need to say something about the outcome equation parameters or does this parameter capture all the parameters in the model and we can delete treatment in the last sentence} \textcolor{red}{It contains parameters in both equations}

\begin{assumption}\label{rank}
(i)$E(\widetilde{Z}_{i}\widetilde{Z}_{i}')$ has full rank
and $E(\widetilde X_{i}\widetilde X_{i}^{\prime })$ has full rank.
\end{assumption}

\begin{assumption}
\label{npl} (i) $\Theta$ is compact, $\theta_0$ is an interior point of $\Theta$, and $\mathcal P_n$ is a compact and convex subset of $(0,1)^n$; (ii) $(\theta_0,P_{[n]}^*)$ is an isolated population NPJL fixed point; i.e., there is an open ball around it that does not contain any other element of $\Lambda_{0n}$;  (iii) the operator $\phi_0(P)-P$ has a nonsingular Jacobian matrix at $P_{[n]}^*$; (iv) there exist non-singular matrices $V_1(\theta_0)$ and $V_2(\theta_0)$ such that
\begin{align*}
\mathbb E\Bigg[\frac{\partial^2 L_n(\theta_0,P_{[n]}^*)}{\partial\theta\partial \theta'}+\frac{\partial^2L_n(\theta_0,P_{[n]}^*)}{\partial \theta\partial P'}\cdot \Bigg[I-\Bigg(\frac{\partial \Psi(\theta_0;P_{[n]}^*)}{\partial P}\Bigg)'\Bigg]^{-1}\cdot\frac{\partial \Psi(\theta_0;P_{[n]}^*)}{\partial \theta'}\Bigg |\mathbb I_O\Bigg]&\xrightarrow{p} \Omega_1(\theta_0),\\
\mathbb E\Bigg[\frac{\partial l_i(\theta_0,P_{[n]}^*)}{\partial\theta}\frac{\partial  l_i(\theta_0,P_{[n]}^*)}{\partial\theta'}\Bigg|\mathbb I_O\Bigg]&\xrightarrow{p} \Omega_2(\theta_0).
\end{align*}
Moreover, $\Omega_1(\theta_0)$ is negative definite.
\end{assumption}

\begin{comment}\begin{assumption}\label{npl}
(i) $\Theta $ is compact, $\theta _{0}$ is an interior point
of $\Theta $, and $\mathcal{P}_n$ is a compact and convex subset of $(0,1)^{n}$%
; (ii) $(\theta _{0},P^{\ast }_{[n]})$ is an isolated population NPJL fixed point;
i.e., it is either unique or there is an open ball around it that does not
contain any other element of $\Lambda _{0}$; (iii) $\frac{\partial ^{2}L_{0}(\theta ,P^{\ast }_{[n]})}{\partial \theta \partial \theta'}$
is a nonsingular matrix; (iv) the operator $\psi _{0}(P_{[n]})-P_{[n]}$ has a
nonsingular Jacobian matrix at $P^{\ast }_{[n]}$; (v) $I-\Big(\frac{%
\partial \Gamma (\theta _{0};P^{\ast }_{[n]})}{\partial P}\Big)'$ is invertible for $n$ sufficiently large; (vi) there exist non-singular
matrices $V_{1}(\theta _{0})$ and $V_{2}(\theta _{0})$ such that: 
\begin{align*}
\mathbb{E}\Bigg[\frac{\partial ^{2}L(\theta _{0},P^*_{[n]})}{%
\partial \theta \partial \theta'}+\frac{\partial ^{2}L(\theta
_{0},P^*_{[n]})}{\partial \theta \partial P'}\cdot \Big[I-\Big(\frac{%
\partial \Gamma (\theta _{0};P^*_{[n]})}{\partial P}\Big)'\Big]^{-1}\cdot 
\frac{\partial \Gamma (\theta _{0};P^*_{[n]})}{\partial \theta '}\Bigg| \mathbb I_T\Bigg]& %
\rightarrow\Omega _{1}(\theta _{0}), \\
\mathbb{E}\Bigg[n\frac{\partial L(\theta _{0},P^*_{[n]})}{\partial
\theta }\frac{\partial L(\theta _{0},P^*_{[n]})}{\partial \theta'}\Bigg|\mathbb I_T\Bigg]& \rightarrow\Omega _{2}(\theta _{0}).
\end{align*}%
Moreover, $\Omega _{1}(\theta _{0})$ is negative definite.
\end{assumption}
\end{comment}

\begin{theorem}
When \Cref{msi,normalization,rank} and \Cref{npl}(i)-(iv) hold, $\hat{\theta}\xrightarrow{p}\theta _{0}$. When additionally, \Cref{npl} (v) also holds, we have:
\begin{equation*}
\sqrt{n}(\hat\theta-\theta_0)\xrightarrow{d}N\left( 0,\Omega
_{1}(\theta_0)\Omega _{2}^{-1}(\theta_0)\Omega _{1}^{\prime
}(\theta_0)\right) .
\end{equation*}
\end{theorem} 
The proof is similar to that of \cite{lin2024binary} and omitted here. Given the  $\sqrt n$ convergent rate of $\hat\theta$, we can show the consistency and convergent rate of $\widehat P_{[n]}$ to its true $P_{[n]}^*$. Following the equilibrium condition, we have: 
  \begin{equation}\label{taylor2}
\Big[I-\Big(\frac{\partial \Psi(\theta^-;P^-)}{\partial P}\Big)'\Big]\big(\hat P_{[n]}-P_{[n]}^*\big)-\frac{\partial \Psi(\theta^-;\pi^-)}{\partial \theta}\big(\hat\theta-\theta_0\big)=0,
\end{equation}
  where $(\theta^-,\pi^-)$ are values between $(\hat\theta,\hat\pi)$ and $(\theta_0,\pi)$. With invertibility of $\Big[I-\Big(\frac{\partial \Psi(\theta^-;P^-)}{\partial P}\Big)'\Big]$, we have: 
  \[
  \|\hat P_{[n]}-P_{[n]}^*\|=O_P\Big(\frac{1}{n}\Big).
  \]
\subsection{Treatment Effects}\label{Model Parameters}

First consider the partial treatment effects captured as the impact of the change of the treatment status of a specific individual holding the treatment status of all other individuals constant. That is, the CCPs in the BNE of the treatment equation are held constant. 
Denote the counterfactual treatment status of individual $i$ as $\overline D_i$. We calculate the average partial treatment effects via the following steps:
\begin{enumerate}
    \item For individual $i$, we calculate the conditional choice probabilities of $Y_i$, for two cases: $\overline D_i=0$ and $\overline D_i=1$, keeping all other $D_{-i}$ fixed. Denote $P^0$ as the CCPs associated with $(\overline D_i=0,D_{-i})$ and $P^1$ as the CCPs associated with $(\overline D_i=1,D_{-i})$.
    \item If $D_i=0$, we have $D=(\overline D_i=0,D_{-i})$ and $P^0=\widehat P_O$. We then use the fixed point algorithm to get $P^1$ based on \Cref{bneo} with $(\overline D_i=1,D_{-i})$.
    \item If $D_i=1$, we have $P^1=\widehat P_O$, and use the fixed point algorithm to get $P^0$ based on \Cref{bneo} with $(\overline D_i=0,D_{-i})$.
    \item Given $P^0$ and $P^1$, the parameter estimates and $(X,Z)$, we obtain $PTE_i$ from \Cref{me}.
    \item Repeating the above steps for $i=1,\cdots,n$, we produce a vector of $PTE_i$ and take their average to obtain the $APTE$.
\end{enumerate}

Now consider the treatment effects induced by changes in the index values in the CCPs. This necessitates evaluating their impact on the values of all of the $D$'s. Although we could obtain new CCPs through the fixed point algorithm, we do not observe the values of the error terms and these are required to generate both the treatment choices and the outcomes. Accordingly, we simulate random shocks from a bivariate normal distribution with the appropriate correlation. With these two error terms and the newly calculated CCPs of treatment and outcome, we can then obtain the counterfactual treatment status and outcomes. We then average over multiple draws of the errors.

This approach is summarized in the following steps: 
\begin{enumerate}
    \item Alter the index value appropriately for each individual. Using the estimated coefficients $(\hat\beta_T,\hat\alpha)$ employ the fixed points calculation to obtain the new CCPs of the treatment choices based on \Cref{bnes}.
    %, i.e., $P(D_i=1|\mathbb I_T),i=1,\cdots,n$.
    \item Simulate the error shocks, $(u_i,v_i)$ from the bivariate normal distribution with means 0, variance 1, and correlation $\hat\rho$, denoted as $u_{ib},v_{ib}, b=1,\cdots, B$ for $B$ simulations.
    \item For each simulation $b$, generate the $D$'s based on $D_{ib}=1\{Z_i'\hat\beta_T+\hat\alpha V^*_i+v_{ib}>0\}$.
    \item With the $D_{ib}'s$ and the estimated $\hat\beta_O,\hat\gamma,\hat \delta$, obtain the CCPs of the outcomes based on \Cref{bneo} using the fixed points calculations.
    %, i.e., $P(Y_i=1|\mathbb I_O), i=1,\cdots,n$. 
    \item Obtain the simulated outcomes as $Y_{ib}=1\{X_i'\hat\beta_O+\hat\gamma D_{ib}+\hat\delta W^*_i+u_{ib}>0\}.$
    \item Repeat the above steps $B$ times to obtain a sequence of $D_b$'s and $Y_b$'s and infer the treatment effects from the respective changes in $D$ and $Y$. 
\end{enumerate}

We explore the performance of these estimation procedures and this is described in \Cref{simulation}. The evidence suggests that they perform well.

\subsection{Selection Models}
In discussing the estimation of selection models we consider two approaches recalling we exclude peer effects from the outcome equation. When the outcome variable is binary the likelihood function presented in \Cref{pll} is applicable with the simplification resulting from the presence of a single Bayesian game. When the outcome is continuous we can adopt two approaches. First, we can simply adjust the likelihood function considered above to incorporate a continuous outcome variable. Second, we can also implement the \cite{heckman1979sample} two step approach in which we estimate the selection equation as in shown in \Cref{selection} after it is simplified to reflect a univariate selection/treatment decision. It is straightforward to then compute the control function from these selection equation estimates which can then be used to include in a second step regression over the appropriately chosen subsample. 
\section{Exercise and Self-Esteem}\label{esteem}

We now investigate the impact of an individual's frequency of exercise on
their level of self-esteem in the presence of peer effects. Self-esteem is considered an important aspect of an individual's
quality of life and mental well-being. For example, \cite{campbell1984new} refers to
self-esteem\ as the "First Law of Human Nature" and improvement in self-esteem is frequently an objective in studies of health
intervention. While many factors are likely to affect an 
individual's self-esteem, empirical evidence
suggests that an individual's level of physical exercise is 
an important determinant  \citep*[see, for example,][]
{sonstroem1984exercise,sonstroem1989exercise,sonstroem1994exercise}. This
is based on existing studies utilizing randomized controlled
trials and/or experiments \citep*[see, for example,][]{ekeland2005can,fox2000self,tiggemann2000effect}. One proposed mechanism is
that exercise affects an individual's sense of
autonomy and personal control over one's physical appearance and functioning
\citep{fox2000effects}. A substantial empirical literature has explored this
relationship \citep*[see, for example,][]{fox2000effects,spence2005effect} and it suggests policies aimed at increasing exercise may increase self-esteem.

One factor which has been ignored in evaluating this relationship is the role of peer effects in both the determination of exercise and self-esteem. An individual's exercise
frequency is likely to be influenced by their expectation of
that of their friends'. This may capture peer pressure, other forms of motivation,
or simply the capacity to participate in exercise which requires multiple
participants. Moreover, recent evidence that one's "feelings" are also subject to contagion \citep*[for example, see][]{juszkiewicz2020self} and suggests a role for peer effects in determining in self-esteem. We contribute to this existing evidence on the impact of exercise on self-esteem by allowing peer effects to determine both.

In addition to incorporating peer effects an 
additional challenge in estimating the impact of
exercise is that it is likely to be endogenous to self-esteem %in that
%same unobservable factors are likely to determine both 
%and reverse causation may exist 
\citep*[see][]{furnham2002body,strelan2003brief}. 
We incorporate these considerations through our procedure described above by estimating the following model: 
\begin{align*}
Self\mbox{-}Esteem_{i}& =1\{X_{i}^{\prime }\beta _{O}+\gamma Exercise_{i}+\delta W_i^*+u_{i}>0\}
\\
Exercise_{i}& =1\{X_{i}'\beta_T+\alpha V_i^*+v_{i}>0\}
\end{align*}%
where $Exercise$ is measured by an indicator function denoting that the
individual's level of weekly exercise is above some specified frequency threshold; $%
Self\mbox{-}Esteem$ is a measure of self-reported self-esteem described
below; the $X^{\prime }s$ denote a vector of exogenous explanatory variables
and the $W^*$ and $V^*$ denote the network variables. The error terms are potentially correlated to allow for the possible endogeneity.
%The primary objects of interest are related to the various treatment effects discussed above. 
%The impact of these peer effects is captured by the parameters $ \delta $ and $
%\alpha $. 
%Peer effects would also capture the contagion phenomena known to exist across a range of activities. We also explore whether similar effects are operating in the determination of self-esteem and examine whether an individual's perception of her friends' level of self esteem directly affect her own. Empirical evidence supports the presence of emotional or self-esteem contagion reflecting the proneness for one to catch the emotions of others.

%As the level of $Exercise$ reflects an individual's choice it is
%likely that it jointly determined with $Self\mbox{-}Esteem$  in that correlated unobserved factors are influencing both. It is
%necessary to accommodate this in estimation. 
We highlight that our private information assumption seems reasonable here. It is likely that individuals do not know how much their friends are exercising. Nor is it likely an individual observes the level of their friends' self-esteem. It is therefore reasonable to assume individuals have only an expectation of each. We include the same exogenous covariates in both equations as the model is identified via the inclusion of the peer effects variable in the treatment equation. However, we also explore the impact on our results of employing different covariates in the two equations. 
%rather than exact measures. 
%This is particularly so given that the exercise variable includes out of school activity. This increases the likelihood that at least some of the exercise activities are not observed by members of the network.

%Following a description of the data, we estimate the model in two ways. As the $Exercise$
%variable is an endogenous binary treatment we first estimate a model where
%the self-esteem variable is treated as a continuous outcome. We then
%consider an empirical model where $Self\mbox{-}Esteem$ is treated as a binary
%outcome denoting the individual is above or below some level on the
%continuous measure. Although each of the models is reliant on the same
%distributional assumptions, we estimate both to illustrate the applicability
%of our approach.
%

\subsection{Data}

We employ data from Wave I of the \textit{National Longitudinal Study of
Adolescent Health} (\textit{Add Health) }dataset which was conducted in
1994-1995 and surveys students in grades 7-12 from a sample of representative
schools. The survey collects information on a number of student
characteristics including academic performance, health conditions, and socio-economic demographic variables.  The survey asks each student to
list as many as five best female and five best male friends. We construct
the friendship network using these responses. We have usable data on 9036 students.

The self-esteem measure is constructed using six items in the 
\textit{Add Health} dataset based on the \cite{rosenberg2015society}
self-esteem scale. Students are asked about their level of
agreement with the following six statements regarding their self-perception
or self-worth: i) \textquotedblleft You have a lot of good
qualities\textquotedblright ; ii) \textquotedblleft You have a lot to be
proud of\textquotedblright ; iii) \textquotedblleft You like yourself just
the way you are\textquotedblright ; iv) \textquotedblleft You feel like you
are doing everything just about right\textquotedblright ; v)
\textquotedblleft You feel socially accepted\textquotedblright ; and vi)
\textquotedblleft You feel loved and wanted\textquotedblright . The
permitted responses are \textquotedblleft strongly agree\textquotedblright ,
\textquotedblleft agree\textquotedblright , \textquotedblleft neither agree
nor disagree\textquotedblright , \textquotedblleft
disagree\textquotedblright , and \textquotedblleft strongly
disagree\textquotedblright\ and each answer is coded 0- 4 with the higher
score indicating greater self-esteem. We add the responses to these statements to produce a self-esteem index which 
ranges from 0 to 24. We create a binary variable indicating the index value is above a specific threshold. This accommodates our inability to estimate models which feature continuous treatments or outcomes. We vary the value of the threshold and estimate a range of models for the probability of being above different quantiles of the index.

The exercise variable is constructed using responses to the question:
\textquotedblleft How many times in a normal week do you work, play, or
exercise hard enough to make you sweat and breathe
heavily?\textquotedblright . The possible responses are: i)
\textquotedblleft never\textquotedblright ; ii) \textquotedblleft 1 or 2
times\textquotedblright ; iii) \textquotedblleft 3 to 5
times\textquotedblright ; iv) \textquotedblleft 6 or 7
times\textquotedblright ; and v) \textquotedblleft more than 7
times\textquotedblright . We define the exercise variable as an
indicator function taking the value 1 if the individual responds either
\textquotedblleft 6 or 7 times\textquotedblright\ or \textquotedblleft more
than 7 times\textquotedblright\ and zero otherwise. We employ this
binary exercise measure as our treatment variable.

As the sample comprises high school students, it would not be surprising if each responded that they did something corresponding to this condition each school day, so the choice of 6 or more times probably corresponds to exercise which includes activities conducted out of school time. Whether this is the correct definition is an empirical question. However, note it is consistent with an important feature of our approach. That is, it is the expectation of peer's exercise, and not the observed level, that is the appropriate variable. As "more than 5 times a week", probably involves exercise out of school time, this increases the likelihood that an individual is basing their own exercise on the expectation of their friends' exercise rather than on how frequently they observe their friends exercising.\footnote{Although this decision to define the treatment as ``more than 5 times a week'' is based on the stated logic, it would be interesting to examine the use of other thresholds. This would then estimate the impact of alternative exercise
treatments on self-esteem. This would also have implications for the construction and interpretation of the peer effects.}
%Although we do not report the results here an initial investigation
%revealed that at low levels of frequency, there is neither a peer effect nor an 
%exercise effect. This suggests that the exercise has a heterogeneous impact that varies across exercise levels both with respect to the peer effect and the impact on self-esteem.}

The assumption of the exogeneity of the friends seems impossible to relax given the large sample. However, it is plausible as individuals may form friendships on the basis of non exercise related activities. However, it could be problematic if friendship is related to joint participation in sporting activities.

\Cref{stat} provides the summary statistics of the variables employed in the 
empirical investigation. A small
majority of the sample is female and over 60 percent are white. The self-esteem variable has a mean of nearly 19, indicating an average response of
"agree" to the various statements. The first
quartile of the distribution of responses is 17 indicating a large fraction
in this sample reporting levels of self-esteem above the mean. However, there
are many individuals reporting a very low value of self-esteem. The median value is 18 and the third quartile is 22.  The exercise
variable has an average value of .4, indicating that 40 percent of the sample
exercises at least 6 times a week. 
%This may appear high, but given the sample it is not surprising.

\subsection{Main Results}

%\textcolor{blue}{is the following possible? I think we should do it. We will be critized for the binary/binary structure....I suggest we define different treatments and ouctomes based on the same indices. For example, each above median, each about 1st quartile, each above 3rd quartile, if we could do each at 1,2, and 3 quartile that would give us 9 treatment effects and 9 sets of parameters..i think this is worth doing} \textcolor{red}{I have results of bottom 10\% quantile in the appendix which has significant peer effects in self-esteem.}

We begin by estimating the self-esteem equation by probit without including either of the network variables nor accounting for the potential endogeniety of exercise. We estimate the model by setting the outcome self-esteem variable as an indicator function that the individual's level of self-esteem is above the median level. The treatment variable is defined as the binary exercise variable. The model also includes a number of
background variables such as the individual's GPA, the individual's age and
gender, family income, and a self-reported intelligence measure.
These results are reported in Table B.1 and they reveal that the impact of exercise is positive and statistically significant and large. The estimated average treatment effect ($APTE$) from exercising is 8.7 percent.  

To test for potential endogeneity of exercise in this restricted model we include the generalized residual from the exercise equation, reported in Table B.2, in the self-esteem equation \citep[see][]{vella1992simple}. These estimates are consistent under the null hypothesis of exogeneity. The t-statistic on the generalized residual is 2.7 indicating that exogeneity can be rejected. As the estimates are inconsistent we cannot interpret their meaning. However, it is notable that the coefficient on the generalized residual is negative and large indicating that the correlation in the unobservables across equations is negative. This is consistent with some correlated unobserved factors negatively influencing exercise and greatly increasing self-esteem. Moreover, the inclusion of the residuals results in a large increase in the exercise variable's coefficient. We delay a discussion of the meaning of coefficients to the model which is identified through the inclusion of the network variables.

%These
%results are reported in ?? and show that a number of the regressors
%have a statistically significant impact on the level of $Self\mbox{-}Esteem$%
%. We delay a discussion of the impact of these other regressors until we
%determine the preferred specification. The estimate of the impact of
%exercise in this model is .733. The estimate is statistically significant
%but surprisingly small in magnitude. Given the construction of $Self\mbox{-}%
%Esteem$, this corresponds to less than the difference in the level of
%response to one sub-question.

%We now estimate the model accounting for the possible endogeneity and the
%presence of peer effects. We estimate the model by joint MLE and the two-step approach, which requires the construction and use of the control
%function. We include the same variables in the exercise equation, noting
%the model is identified by the inclusion of the peer effects variable, which
%is a function of the $X^{\prime }s$ of the other members in the network. We
%estimate the $Exercise$ equation both separately and jointly with the $Self%
%\mbox{-}Esteem$ equation. We only discuss the results for the joint MLE
%estimation of the $Exercise$ equation, in \Cref{selectionmle}, noting that the
%estimates for the separate estimation, reported in \Cref{selection}, are almost
%identical.

\begin{table}[h]
\caption{Statistic Summary of Variables}
\label{stat}\centering
\begin{tabular}{rcc}
\hline\hline
Variable & Mean & Standard Deviation \\ \hline
Age & 15.638 & 1.635 \\ 
Female & 0.525 & 0.518 \\ 
GPA & 2.834 & 0.797 \\ 
Intelligence (1\&2) & 0.048 & 0.214 \\ 
Intelligence (3\&4) & 0.589 & 0.492 \\ 
Intelligence (5\&6) & 0.363 & 0.481 \\ 
White & 0.619 & 0.486 \\ 
Family Income & 0.049 & 0.055 \\ 
Exercise & 0.408 & 0.491 \\ 
Self-Esteem & 0.497 & 0.500 \\ \hline\hline
\end{tabular}%
\end{table}

\begin{table}[ht!]
\caption{Joint MLE Estimation of Exercise and Self-Esteem}\label{main}
\centering
\begin{tabular}{rcc}
\hline\hline
\multicolumn{3}{c}{Exercise}\\\hline
Variable & Estimate & Standard Error \\ \hline
Age & -0.106*** & 0.009 \\ 
Female & -0.668*** & 0.027 \\ 
GPA & 0.049*** & 0.018 \\ 
Intelligence (3\&4) & 0.024 & 0.064 \\ 
Intelligence (5\&6) & 0.085 & 0.067 \\ 
White & 0.224*** & 0.029 \\ 
Family Income & 0.659*** & 0.250 \\ 
Intercept & 1.378*** & 0.158 \\ \hline
Peer Effects & 0.157*** & 0.061 \\ \hline
\multicolumn{3}{c}{Self-Esteem}\\\hline
Age & 0.004 & 0.015 \\ 
Female & 0.088 & 0.078 \\ 
GPA & -0.005 & 0.018 \\ 
Intelligence (3\&4) & 0.143** & 0.063 \\ 
Intelligence (5\&6) & 0.475*** & 0.081 \\ 
White & -0.171*** & 0.030 \\ 
Family Income & -0.233 & 0.242 \\ 
Exercise & 1.296*** & 0.195 \\ 
Intercept & -0.763** & 0.292 \\ \hline
Peer Effects& -0.020&0.050\\
$\rho$ & -0.695** & (-0.894,-0.334) \\ \hline\hline
\multicolumn{3}{l}{For $\rho$, profiled confidence interval is reported.}\\
\multicolumn{3}{l}{*** 1\%; ** 5\%; * 10\%}\
\end{tabular}%
\end{table}

\begin{table}[ht!]
\caption{Joint MLE Estimation of Exercise and Self-Esteem(Bottom 10\%)}\label{bottom10}
\centering
\begin{tabular}{rcc}
\hline\hline
\multicolumn{3}{c}{Exercise}\\\hline
Variable & Estimate & Standard Error \\ \hline
Age & -0.107*** & 0.009 \\ 
Female & -0.665*** & 0.027 \\ 
GPA & 0.050*** & 0.019 \\ 
Intelligence (3\&4) & 0.028 & 0.065 \\ 
Intelligence (5\&6) & 0.091 & 0.068 \\ 
White & 0.229*** & 0.029 \\ 
Family Income & 0.672*** & 0.250 \\ 
Intercept & 1.372*** & 0.159 \\ \hline
Peer Effects & 0.154** & 0.062 \\ \hline
\multicolumn{3}{c}{Self-Esteem}\\\hline
Age & -0.003 & 0.019 \\ 
Female & -0.141 & 0.108 \\ 
GPA & 0.064** & 0.025 \\ 
Intelligence (3\&4) & 0.282*** & 0.075 \\ 
Intelligence (5\&6) & 0.460*** & 0.086 \\ 
White & -0.118*** & 0.045 \\ 
Family Income & -0.161 & 0.352 \\ 
Exercise & 0.830*** & 0.321 \\ 
Intercept & 0.536 & 0.455 \\ \hline
Peer Effects& 0.075*&0.041\\
$\rho$ & -0.251 & (-0.491,0.017) \\ \hline\hline
%\multicolumn{3}{l}{Peer Effects parameter significance is based on one-sided test.}%
\end{tabular}%
\end{table}

We now estimate the full model and the results are reported in Table 2. A number of features of the estimates from the $Exercise$ equation are of
interest, as several of the explanatory variables are statistically
significant. Exercise decreases with age, although there is not a great deal
of variation in age across individuals in our sample. Females exercise less
and students who respond that they are white exercise more. The estimated female
effect seems large. Higher GPA and family incomes increase the
probability that the individual exercises more than 5 times a week. The most interesting coefficient is associated with the expected behavior of the individual's peers. The estimated coefficient is .157 and it is statistically significant at the 5 percent level. As the standard deviation, the minimum and maximum of this variable are 0.225, 0 and 0.768 respectively, the impact on the probability of exercising more than 5 times a week is not small. The marginal impact of going from the lowest to the highest value of $V$ is to increase the average probability of exercise from .396 to .440. This is not surprising given the various different ways such an effect is likely to occur. The magnitude of this marginal effect is consistent with our assumption 2 regarding the level of social influence.

Now focus on the estimates from the $Self\mbox{-}Esteem$ equation. A number of the background variables have statistically significant impacts on the individual's level of self-esteem. The students who report having higher intelligence have higher self-esteem while white individuals report lower self-esteem. The treatment variable is large in magnitude and highly statistically significant. The coefficient suggests a large positive effect on self-esteem from exercise. In contrast to the exercise equation, there is no evidence of peer effects operating in the self-esteem equation. The estimate of $\rho$ is negative and statistically different from zero. The
negative correlation between the unobservables appears reasonable, although the direction of this
relationship appears to be an empirical matter.  We estimate an average partial treatment effect, shown in equation (\ref{ate}), of 0.543.

%In addition to the magnitude of the $Exercise$ coefficient, there are other
%notable differences across the two sets of estimates. The most remarkable
%difference is the larger and statistically significant effect for the female
%variable. The estimates from the joint MLE procedure are generally more
%similar to the unadjusted OLS estimates although the exercise effect is much
%larger for the joint MLE estimates. While we do not test it formally, it is
%possible that the two adjusted estimates are not statistically different.
%Although it is beyond the scope of this paper to investigate this issue, it
%is possible that the joint MLE procedure's heavier reliance on normality is
%responsible for the difference across the estimates.

The large increase in the $APTE$ relative to the earlier results requires explanation. One interpretation is that it is due to identification via the inclusion of the peer effects in the exercise equation. Accordingly, we excluded family income from the self-esteem equation and re-estimated with and without the peer effects included. These results are reported in 
%the appendix in \Cref{exclusionpeer} and \Cref{exclusion} 
Tables B3 and B4 respectively. In both cases, the coefficients on $Exercise$ and $\rho$ were very similar to those in \Cref{main}. These results suggest that the large increase in the treatment effect is not due to the identification approach. Rather, it reflects the substantial role operating through the endogeneity. This is confirmed by the estimate of the $APTE$ of .544 for the model identified via the exclusion restriction.

It is important to consider why accounting for endogeneity operating through the treatment decision leads to such a large increase in the treatment effect. The large estimated value of $\rho$ suggests that the correlation in the errors across equations is very high and negative. This indicates that there are potentially some important common unobserved determinants  across equations but that the sign of their influence differs across equations. These common unobserved factors might reflect the relatively small number of observed characteristics available in the data set which influence exercise and self-esteem. For example, the variables employed here are those employed in more typical economic investigations. However, perhaps the factors which explain exercise and self-esteem are not collected or are unobserved. To the extent these "uncontrolled factors" are correlated this will result in a high correlation and a potentially large effect on the treatment coefficient. It is not clear why the correlation across equation is negative other than it is also likely to capture the complicated nature of the determination of self-esteem. However, it suggests that there are unobserved factors which are simultaneously reducing (increasing) self-esteem while increasing (reducing) exercise. Moreover, this strong negative correlation is masking the large positive impact of exercise on self-esteem. While we interpret our results as clearly supporting the positive impact of exercise on self-esteem we acknowledge this treatment effect is likely to vary with additional conditioning variables. Note that while our results are consistent with earlier empirical work regarding the sign of the relationship between self-esteem and exercise, we cannot make direct comparisons with this work regarding the magnitude of the effect as earlier studies focus solely on its sign. Our results also clearly support the presence of peer effects in the exercise equation.

Before we analyze the economic significance of these results in terms of outcome responses to targeted interventions, we explore the presence of peer effects at different deciles of the self-esteem index variable. We re-estimate the model for outcomes reflecting that the self-esteem variable is above different deciles of the index. The results were generally similar to those for the median with one striking exception. In \Cref{bottom10} we report the results for the individual's value of self-esteem being above the bottom decile. Given the lumpiness of the index, reflected in Table B.6, this corresponded to the 10.8 quantile. A number of the results are worth noting. First, the exercise coefficient has diminished relative to the results when the outcome is based on median self-esteem but remains large. Second, the estimate of the $\rho$ is smaller in absolute terms and there is less evidence of statistical significance. Finally, the peer effects variable in the self-esteem equation is positive and statistically significant at the 10 percent level. This is an interesting result as it suggests that there is contagion for those with lower levels of self-esteem. This is consistent with existing evidence. We estimated the model at higher deciles and several patterns arose. These are reported in Tables B.7 to B.11 noting that due to nature of the index shown in Table B.6 unique results do not exist at each decile.  First, the statistical significance of the peer effects in the self-esteem equation decreased as the decile examined increased. The exercise coefficient is relatively constant across deciles with the exception of the top and bottom. Finally, the estimate of $\rho$ varied by decile and was sometimes not statistically different from zero. This suggests the unobservables play different roles at different points in the distribution of the self-esteem index. The results for the other quantiles are reported in \Cref{tables}.

\subsection{Treatment Effects and Counterfactual Studies}\label{counterfactual}
We now consider the implications of directly influencing the treatment decisions for two subsamples of the data. We evaluate how doing so affects the treatment choices and the level of the entire sample's self-esteem. In both instances, we increase the value of the single index in the treatment decision.  We first select the 1,000 students with the lowest family incomes and increase their single index by 0.5 noting that the standard deviation of this index is .409. We then calculate the conditional choice probabilities of the Bayesian game of treatment choices. By employing simulated $v_i$'s, we can generate the associated $D_i's$. We then calculate the CCPs of outcomes using the estimated parameters and the generated $D$'s based on \Cref{bneo}. Using the CCPs for the outcome index, and the $u_i's$  simulated according to the estimated covariance information, we obtain the self-esteem status of all students. In the second counterfactual analysis, we simulate a .5 increase in the index value for students based on their level of popularity in the network. We use the in-links information to calculate each individual's total friends nominations from others. Using this information we infer their level of popularity and select the 1,000 most popular. We then employ the same process as above. For both of these interventions we conduct these processes 1,000 times and report the average results.

First, consider the counterfactual for the 1000 poorest students. Table 4 reports that the APTE for these students is .626 which is notably higher than the sample value of .544. This reflects the heterogeneity in the treatment effect. To better understand the ongoing mechanisms, we first examine the behavior of the 1000 selected students before and after the intervention. Prior to the index increase, 357 students are exercising and 494 reported above median self-esteem. These increase to 538 and 585 respectively with the change in the index. This corresponds to a composite treatment effect for these targeted individuals of (585-494)/(538-357)= 50.2\%. Note that this is not comparable to the APTE as we are not holding the indices constant and we do not change each individuals' value of treatment from 0 to 1. Nevertheless, it does indicate how these individuals change their behavior. It is somewhat unexpected that the value of this composite treatment effect is lower than the corresponding ATE of .626. % However, it reflects the heterogeneity in the model. 
Now consider the effect on the entire sample. Table 5 reports that the number of individuals reporting an increase in exercise increases by 188. This is only slightly higher than the 181 in Table 4 indicating the presence of peer effects in the exercise decision are increasing the numbers increasing by 3.8\%. The increase in those reporting above median self-esteem is 111. This is an increase of 22\% over those in the Table 4. This illustrates the strength of equilibrium and peer effects. The composite treatment effect for the whole sample is  $(4600-4489)/(3872-3684)=59.0\%$.

Now consider the second counterfactual in which the indices for the 1000 most popular students are increased. The APTE for this group of individuals is .563 and their composite treatment effect is  $(614-529)/(614-476)=61.6\%$. The impact of this intervention for the entire sample is $(4608-4489)/(3883-3684)=59.8\%$.  The peer effects are also important in this intervention. For example, the increase in exercise for the targeted group is 138 (i.e. 614-476) while the change in the whole sample is 199 (i.e. 3883-3684). This is an increase of 44\%. The change in self-esteem for the targeted group is 85 (i.e. 614-529) while for the whole sample, it is 119. This is a 27\% increase. This is strongly suggestive of peer effects affecting both the treatment and outcome equations recalling that peer effects in even only the treatment equations have implications for the observed outcomes. It also highlights the targeting of the appropriate group of individuals via policy in the presence of peer effects. 

These two examples illustrate the importance of peer effects in this setting. In both instances, individuals beyond the targeted group are changing their exercise decision as a result of a change in the targeted group's behavior. This is not surprising given the evidence of peer effects in the exercise decision model. There is also evidence that the impact on self-esteem is also going beyond the targeted group. This is capturing both the direct effect of changing the exercise decision of this group and the peer effects in the self-esteem equation. The examples also illustrate the potential importance of identifying the appropriate targeted group when the sole criteria is maximizing the number of people whose outcome is affected. While the differences are not large, it appears that in this setting the more effective policy, using this simple criterion, is to influence the treatment decision of the most popular students.

%Table 7 reports the outcomes from these two counterfactuals. In the first experiment the number of students increasing their level of exercise increases by 181 which represents an increase of 50.7 percent. The number increasing their level of self-esteem is 91 which is a change of 18.4 percent. 
%We have to see how much of the change is due to the 1,000 changing self-esteem and how much is due to the network effects. This represents a treatment effect of 50.3 percent. In the second experiment the increase in the exercise index of the most popular leads to an increase of 138 individuals changing their exercising behavior. This is a change of 29.0 percent. This results in a change in the number with higher self-esteem of 85 (16.6 percent) which represents a treatment effect of 61.6 percent. The calculated AME of the first scenario is 56.3 percent.

%\textcolor{blue}{what can we say about direct and indirect effects? what can we say about the magnitude of this effect and the average treatment effect?}

\begin{table}[ht!]
\caption{Counterfactual Decisions and Outcomes}\label{binaryselection}
\centering
\begin{tabular}{c|c|c}
\hline\hline
\multicolumn{3}{c}{Counterfactual 1}\\\hline
 & Observed & After Index Change \\ \hline
Exercise&357& 538***(50.7\%+)\\
&&(16.318)\\
High Self-Esteem&494&585***(18.4\%+)\\
&&(15.053)\\\hline
APTE&\multicolumn{2}{c}{0.626}\\\hline
\multicolumn{3}{c}{Counterfactual 2}\\\hline
 & Observed & After Index Change \\ \hline
Exercise&476& 614***(29.0\%+)\\
&&(15.131)\\
High Self-Esteem&529&614***(16.1\%+)\\
&&(16.273)\\\hline
APTE&\multicolumn{2}{c}{0.563}\\
\hline\hline
\end{tabular}%
\end{table}

\begin{table}[ht!]
\caption{Counterfactual Decisions and Outcomes for whole sample}\label{WholeC}
\centering
\begin{tabular}{c|c|c|c}
\hline\hline
 & Observed & Counterfactual 1& Counterfactual 2 \\ \hline
Exercise&3,684& 3,872***(5.1\%+)&3,883***(5.4\%+)\\
&&(44.286)&(41.646)\\
High Self-Esteem&4,489&4,600**(2.5\%+)&4,608***(2.7\%+)\\
&&(48.852)&(45.301)\\
\hline\hline
\end{tabular}%
\end{table}
%In \Cref{WholeC}, we have ATEs of the first and the second counterfactuals for the whole sample as $(4600-4489)/(3872-3684)=59.0\%$ and $(4608-4489)/(3883-3684)=59.8\%$, compared to $(585-494)/(538-357)=50.3\%$ and $(614-529)/(614-476)=61.6\%$.

\section{Conclusion}\label{conclusion}
We provide a methodology to estimate endogenous treatment models with social interactions by incorporating a role for
an individual's peers' treatment and outcome choices in determining their own. This is done using a game-theoretic approach based on 
discrete Bayesian games which capture the simultaneity of choices. Accounting for this interaction
between choices introduces a computational burden in estimation which we
address via a nested pseudo joint likelihood estimator.  We also describe how our approach extends to the estimation of parameters in
sample selection models. We implement our procedure to examine the impact of an individual's level
of exercise on their self-esteem in the presence of social interactions and the potential endogeneity of exercise. We find that an individual's level of
exercise is influenced by their expectation of their peer's exercise
activity. Accounting for its endogeneity leads to exercise having a large and statistically significant impact on an individual's self-esteem. We also find evidence of peer effects at lower levels of self-esteem. We also find via an examination of the impact of targeted interventions that the peer effects are economically important. 
Finally, while our empirical focus is a health/labor economics example, our methodology can also be employed in industrial organization settings to study, for example, triangular relationships in markets. This would include competitive entry decisions and strategic promotion decisions of firms. For both the entry and promotion decisions, there are likely to be strategic interactions among firms/brands in the market.

\end{doublespace}

\phantomsection
\addcontentsline{toc}{chapter}{References}
\bibliographystyle{chicago}
\bibliography{selection}

\newpage
\crefalias{section}{appsec}

\begin{appendices}
\section{Proof of Theorem 1}\label{appendix A}
\begin{proof}[Proof of \Cref{unique}] We prove the uniqueness of both \Cref{bnes} and \Cref{bneo}. The existence is guaranteed by the Brouwer fixed point theorem. For the Bayesian game of treatment choice, start with any $P,P'\in [0,1]^n$ and let $\tilde P=\Gamma(P)$ and $\tilde P'=\Gamma(P')$. For every $i\in\mathcal I$, it follows that:
\begin{align*}
|\tilde P_{i}-\tilde P_{i}'| &=|\Phi[Z_i'\beta_D+\alpha V_i(P)]-\Phi[%
Z_i'\beta_D+\alpha V_i(P')]| \\
&=\phi[Z_i'\beta_D+ \alpha V_i(\bar P)]| \cdot\Big|\alpha\frac{1}{N_{i}}%
\sum_{j\in F_{i}}(P_{j}-P_{j}')\Big| \\
&\leq \frac{1}{\sqrt{2\pi}}\alpha\cdot \max_{j\in F_{i}}|P_{j}-P_{j}'|
\\
&<\max_{j\in \mathcal{I}}|P_{j}-P_{j}'|
\end{align*}%
where $\bar{P}=(\bar{P}_{1},\cdots ,\bar{P}_{n})$ is
between $P$ and $P'$. Maximizing over $\mathcal{I}$ on the
left-hand side, we obtain:

\begin{equation*}
\max_{i\in \mathcal{I}}|\tilde P_{i}-\tilde P_{i}'|<\max_{j\in \mathcal{I}%
}|P_{j}-P_{j}'|
\end{equation*}%
which is a contraction. By the contraction mapping theorem, there exists a unique fixed point for $\Gamma$ in $[0,1]^n$, say $P^*$.
So $P^*=\Gamma(P^*)$. Since $P=\Gamma (P)$ if and only if $P$ describes an equilibrium, $P^*$ is the unique equilibrium of the Bayesian game. The proof takes a very conservative expansion of terms and the violation of the MSI assumption does not necessarily cause
multiple equilibria.

    For the Bayesian game of outcomes, suppose there are two distinct BNEs, $\bar P\neq \bar P'$.
    \begin{align*}
        |\bar P_i-\bar P_i'|&=|\Psi_i(\bar P;\theta)-\Psi_i(\bar P';\theta)|=\Big|\psi_i[W_i(\widetilde P);\theta]\cdot\delta\cdot\frac{1}{N_i}\sum_{j\in F_i}(\bar P_j-\bar P_j')\Big|,\\&\leq \sup\psi_i(\cdot)\cdot|\delta|\max_{j\in I}|\bar P_j-\bar P_j'|<\max_{j\in I}|\bar P_j-\bar P_j'|,
    \end{align*} where $\widetilde P$ is between $\bar P$ and $\bar P'$. Thus $\max_{i\in I} |\bar P_i-\bar P_i'|<\max_{j\in I}|\bar P_j-\bar P_j'|$ which is a contradiction. Thus, we show the uniqueness of the BNE.
\end{proof}

\newpage
\section{Tables}
\label{tables}
\captionsetup{labelformat=AppendixTables}
\setcounter{table}{0}

\begin{table}[!ht]
\caption{Probit Estimates of Self-Esteem Equation without Peer Effects}\label{selfesteem}
\centering
\begin{tabular}{rcc}
\hline\hline
Variable & Estimate & Standard Error \\ \hline
Age & -0.048*** & 0.008 \\ 
Female & -0.221*** & 0.028 \\ 
GPA & 0.018 & 0.018 \\ 
Intelligence (3\&4) & 0.187*** & 0.064 \\ 
Intelligence (5\&6) & 0.610*** & 0.067 \\ 
White & -0.090*** & 0.028 \\ 
Family Income & 0.052 & 0.247 \\ 
Exercise & 0.227*** & 0.029 \\ 
Intercept & 0.434*** & 0.156 \\
 \hline\hline
\end{tabular}%
\end{table}

\begin{table}[ht]
\caption{Probit Estimates of Exercise without Peer Effects}\label{exercise}
\centering
\begin{tabular}{rcc}
\hline\hline
Variable & Estimate & Standard Error \\ \hline
Age & -0.110*** & 0.008 \\ 
Female & -0.666*** & 0.028 \\ 
GPA & 0.051*** & 0.019 \\ 
Intelligence (3\&4) & 0.033 & 0.066 \\ 
Intelligence (5\&6) & 0.097 & 0.068 \\ 
White & 0.237*** & 0.029 \\ 
Family Income & 0.680*** & 0.253 \\ 
Intercept & 1.436*** & 0.157 \\ 
 \hline\hline
\end{tabular}%
\end{table}

\begin{table}[ht]
\caption{Bivariate Probit Estimates of Self-Esteem and Exercise}\label{bivariate}
\centering
\begin{tabular}{rcc}
\hline\hline
\multicolumn{3}{c}{Exercise}\\\hline
Variable & Estimate & Standard Error \\ \hline
Age & -0.110*** & 0.008 \\ 
Female & -0.669*** & 0.027 \\ 
GPA & 0.052*** & 0.018 \\ 
Intelligence (3\&4) & 0.028 & 0.064 \\ 
Intelligence (5\&6) & 0.090 & 0.067 \\ 
White & 0.233*** & 0.029 \\ 
Family Income & 0.668*** & 0.250 \\ 
Intercept & 1.440*** & 0.156 \\ \hline
\multicolumn{3}{c}{Self-Esteem}\\\hline
Age & 0.003 & 0.015 \\ 
Female & 0.083 & 0.081 \\ 
GPA & -0.005 & 0.018 \\ 
Intelligence (3\&4) & 0.143** & 0.063 \\ 
Intelligence (5\&6) & 0.478*** & 0.082 \\ 
White & -0.171*** & 0.030 \\ 
Family Income & -0.230 & 0.243 \\ 
Exercise&1.284***&0.206\\
Intercept & -0.753** & 0.306 \\\hline
$\rho$&-0.686**&(-0.877,-0.298)\\
\hline\hline
\end{tabular}%
\end{table}

\begin{table}[ht!]
\caption{Exercise and Self-Esteem (Income as Exclusion)}\label{exclusionpeer}
\centering
\begin{tabular}{rcc}
\hline\hline
\multicolumn{3}{c}{Exercise}\\\hline
Variable & Estimate & Standard Error \\ \hline
Age & -0.107*** & 0.009 \\ 
Female & -0.668*** & 0.027 \\ 
GPA & 0.049*** & 0.018 \\ 
Intelligence (3\&4) & 0.025 & 0.065 \\ 
Intelligence (5\&6) & 0.087 & 0.067 \\ 
White & 0.226*** & 0.029 \\ 
Family Income & 0.568** & 0.238 \\ 
Intercept & 1.380*** & 0.158 \\ \hline
Peer Effects & 0.156** & 0.061 \\ \hline
\multicolumn{3}{c}{Self-Esteem}\\\hline
Age & 0.000 & 0.016 \\ 
Female & 0.065 & 0.085 \\ 
GPA & -0.004 & 0.018 \\ 
Intelligence (3\&4) & 0.146** & 0.064 \\ 
Intelligence (5\&6) & 0.488*** & 0.083 \\ 
White & -0.170*** & 0.031 \\ 
Exercise & 1.234*** & 0.226 \\ 
Intercept & -0.686** & 0.320 \\ \hline
Peer Effects& -0.016&0.051\\
$\rho$ & -0.650** & (-0.857,-0.208) \\ \hline\hline
\end{tabular}%
\end{table}

\begin{table}[ht]
\caption{Bivariate Probit Regression of Self-Esteem and Exercise (Income as Exclusion)}\label{exclusion}
\centering
\begin{tabular}{rcc}
\hline\hline
\multicolumn{3}{c}{Exercise}\\\hline
Variable & Estimate & Standard Error \\ \hline
Age & -0.110*** & 0.008 \\ 
Female & -0.669*** & 0.027 \\ 
GPA & 0.052*** & 0.018 \\ 
Intelligence (3\&4) & 0.030 & 0.065 \\ 
Intelligence (5\&6) & 0.093 & 0.067 \\ 
White & 0.235*** & 0.029 \\ 
Family Income & 0.581** & 0.239 \\ 
Intercept & 1.442*** & 0.156 \\ \hline
\multicolumn{3}{c}{Self-Esteem}\\\hline
Age & 0.001 & 0.016 \\ 
Female & 0.059 & 0.088 \\ 
GPA & -0.004 & 0.018 \\ 
Intelligence (3\&4) & 0.147** & 0.064 \\ 
Intelligence (5\&6) & 0.491*** & 0.085 \\ 
White & -0.169*** & 0.032 \\ 
Exercise&1.218***&0.238\\
Intercept & -0.671** & 0.335 \\\hline
$\rho$&-0.638**&(-0.850,-0.221)\\
 \hline
 APTE&0.544&\\\hline\hline
\end{tabular}%
\end{table}

\begin{table}[!ht]
    \centering
    \begin{tabular}{l|c}
    \hline\hline
    Threshold&Percentage reporting low self-esteem\\\hline
         Quantile (0.1)& 10.8\%\\
         Quantile (0.2)& 23.6\%\\
         Quantile (0.3)&33.2\%\\
         Quantile (0.4)&50.3\%\\
         Median&50.3\%\\
         Quantile (0.6)&66.3\%\\
         Quantile (0.7)&73.9\%\\
         Quantile (0.8)&81.3\%\\
         Quantile (0.9) &100\%\\
         \hline\hline
         \multicolumn{2}{l}{The table entries denote the percentage of individuals who are below specific  }\\ 
         \multicolumn{2}{l}{quantiles of the self-esteem index. For example, at the quantile (0.1),}\\
         \multicolumn{2}{l}{10.8\% of sample report low self-esteem (i.e. $Y=0$).}
    \end{tabular}
    \caption{Percentage Reporting Low Self-Esteem with Different Cutoffs}
    \label{tab:my_label}
\end{table}

\begin{table}[ht!]
\caption{Exercise and Self-Esteem Equations (Bottom 20\%)}\label{binaryselection}
\centering
\begin{tabular}{rcc}
\hline\hline
\multicolumn{3}{c}{Exercise}\\\hline
Variable & Estimate & Standard Error \\ \hline
Age & -0.107*** & 0.009 \\ 
Female & -0.668*** & 0.027 \\ 
GPA & 0.051*** & 0.019 \\ 
Intelligence (3\&4) & 0.034 & 0.065 \\ 
Intelligence (5\&6) & 0.095 & 0.068 \\ 
White & 0.229*** & 0.029 \\ 
Family Income & 0.659*** & 0.250 \\ 
Intercept & 1.361*** & 0.159 \\ \hline
Peer Effects & 0.160*** & 0.061 \\ \hline
\multicolumn{3}{c}{Self-Esteem}\\\hline
Variable & Estimate & Standard Error \\ \hline
Age & 0.018 & 0.013 \\ 
Female & 0.011 & 0.075 \\ 
GPA & 0.048*** & 0.020 \\ 
Intelligence (3\&4) & 0.172*** & 0.064 \\ 
Intelligence (5\&6) & 0.365*** & 0.074 \\ 
White & -0.131*** & 0.032 \\ 
Family Income & -0.296 & 0.264 \\ 
Exercise & 1.225*** & 0.172 \\ 
Intercept & -0.429 & 0.292 \\ \hline
Peer Effects& 0.017&0.037\\
$\rho$ & -0.666** & (-0.839,-0.374) \\ \hline\hline
\end{tabular}%
\end{table}

\begin{table}[ht!]
\caption{Exercise and Self-Esteem Equations (Bottom 30\%)}\label{binaryselection}
\centering
\begin{tabular}{rcc}
\hline\hline
\multicolumn{3}{c}{Exercise}\\\hline
Variable & Estimate & Standard Error \\ \hline
Age & -0.106*** & 0.009 \\ 
Female & -0.667*** & 0.027 \\ 
GPA & 0.052*** & 0.018 \\ 
Intelligence (3\&4) & 0.030 & 0.065 \\ 
Intelligence (5\&6) & 0.089 & 0.068 \\ 
White & 0.225*** & 0.029 \\ 
Family Income & 0.677*** & 0.250 \\ 
Intercept & 1.356*** & 0.158 \\ \hline
Peer Effects & 0.161*** & 0.061 \\ \hline
\multicolumn{3}{c}{Self-Esteem}\\\hline
Variable & Estimate & Standard Error \\ \hline
Age & 0.017 & 0.013 \\ 
Female & 0.104 & 0.068 \\ 
GPA & 0.027 & 0.018 \\ 
Intelligence (3\&4) & 0.103* & 0.060 \\ 
Intelligence (5\&6) & 0.328*** & 0.072 \\ 
White & -0.155*** & 0.029 \\ 
Family Income & -0.319 & 0.244 \\ 
Exercise & 1.382*** & 0.132 \\ 
Intercept & -0.662** & 0.261 \\ \hline
Peer Effects& -0.014&0.038\\
$\rho$ & -0.779** & (-0.923,-0.481) \\ \hline\hline
\end{tabular}%
\end{table}

\begin{table}[ht!]
\caption{Exercise and Self-Esteem Equations (Top 40\%)}\label{binaryselection}
\centering
\begin{tabular}{rcc}
\hline\hline
\multicolumn{3}{c}{Exercise}\\\hline
Variable & Estimate & Standard Error \\ \hline
Age & -0.107*** & 0.009 \\ 
Female & -0.667*** & 0.027 \\ 
GPA & 0.049*** & 0.018 \\ 
Intelligence (3\&4) & 0.029 & 0.065 \\ 
Intelligence (5\&6) & 0.092 & 0.067 \\ 
White & 0.228*** & 0.029 \\ 
Family Income & 0.657*** & 0.250 \\ 
Intercept & 1.376*** & 0.158 \\ \hline
Peer Effects & 0.154** & 0.061 \\ \hline
\multicolumn{3}{c}{Self-Esteem}\\\hline
Variable & Estimate & Standard Error \\ \hline
Age & 0.004 & 0.013 \\ 
Female & 0.061 & 0.064 \\ 
GPA & -0.031 & 0.018 \\ 
Intelligence (3\&4) & 0.191*** & 0.069 \\ 
Intelligence (5\&6) & 0.583*** & 0.081 \\ 
White & -0.142*** & 0.031 \\ 
Family Income & -0.209 & 0.248 \\ 
Exercise & 1.174*** & 0.190 \\ 
Intercept & -1.094*** & 0.247 \\ \hline
Peer Effects& -0.044&0.074\\
$\rho$ & -0.595** & (-0.816,-0.272) \\ \hline\hline
\end{tabular}%
\end{table}

\begin{table}[ht!]
\caption{Exercise and Self-Esteem Equations (Top 30\%)}\label{binaryselection}
\centering
\begin{tabular}{rcc}
\hline\hline
\multicolumn{3}{c}{Exercise}\\\hline
Variable & Estimate & Standard Error \\ \hline
Age & -0.107*** & 0.009 \\ 
Female & -0.665*** & 0.027 \\ 
GPA & 0.048** & 0.019 \\ 
Intelligence (3\&4) & 0.031 & 0.065 \\ 
Intelligence (5\&6) & 0.093 & 0.068 \\ 
White & 0.229*** & 0.029 \\ 
Family Income & 0.670*** & 0.251 \\ 
Intercept & 1.376*** & 0.159 \\ \hline
Peer Effects & 0.153** & 0.062 \\ \hline
\multicolumn{3}{c}{Self-Esteem}\\\hline
Variable & Estimate & Standard Error \\ \hline
Age & -0.011 & 0.020 \\ 
Female & -0.048 & 0.113 \\ 
GPA & -0.025 & 0.020\\ 
Intelligence (3\&4) & 0.211*** & 0.076 \\ 
Intelligence (5\&6) & 0.610*** & 0.089 \\ 
White & -0.129*** & 0.042 \\ 
Family Income & -0.115 & 0.278 \\ 
Exercise & 0.753* & 0.409 \\ 
Intercept & -0.940** & 0.428 \\ \hline
Peer Effects& -0.015&0.105\\
$\rho$ & -0.307 & (-0.721,0.190) \\ \hline\hline
\end{tabular}%
\end{table}

\begin{table}[ht!]
\caption{Exercise and Self-Esteem Equations (Top 20\%)}\label{top20}
\centering
\begin{tabular}{rcc}
\hline\hline
\multicolumn{3}{c}{Exercise}\\\hline
Variable & Estimate & Standard Error \\ \hline
Age & -0.107*** & 0.009 \\ 
Female & -0.665*** & 0.028 \\ 
GPA & 0.049*** & 0.019 \\ 
Intelligence (3\&4) & 0.030 & 0.065 \\ 
Intelligence (5\&6) & 0.093 & 0.068 \\ 
White & 0.228*** & 0.029 \\ 
Family Income & 0.667*** & 0.251 \\ 
Intercept & 1.372*** & 0.159 \\ \hline
Peer Effects & 0.156** & 0.062 \\ \hline
\multicolumn{3}{c}{Self-Esteem}\\\hline
Variable & Estimate & Standard Error \\ \hline
Age & -0.022 & 0.024 \\ 
Female & -0.133 & 0.143 \\ 
GPA & -0.024 & 0.023\\ 
Intelligence (3\&4) & 0.208** & 0.085 \\ 
Intelligence (5\&6) & 0.609*** & 0.093 \\ 
White & -0.056 & 0.055 \\ 
Family Income & -0.214 & 0.314 \\ 
Exercise & 0.471 & 0.548 \\ 
Intercept & -0.932 & 0.565 \\ \hline
Peer Effects& 0.014&0.160\\
$\rho$ & -0.130 & (-0.711,0.565) \\ \hline\hline
\end{tabular}%
\end{table}

\begin{doublespace}
\section{Simulation Evidence}\label{simulation}
\captionsetup{labelformat=AppendixTable}
\setcounter{table}{0}
%We now examine the performance of the proposed both joint MLE and two-step estimators via simulation exercises.  

%\subsection{Peer Effects in Treatments and Outcomes}
We generate the data from the following model:
\begin{align}
    \begin{split}
        Y_i&=1\{\beta_{0O}+\beta_{1O}X_{1i}+\gamma D_i+\delta W_i+u_i>0\}\\
        D_i&=1\{\beta_{S0}+\beta_{S1}X_{1i}+\beta_{S1}X_{2i}+\alpha V_i+v_i>0\}
    \end{split}
\end{align}
where $W_i=\frac{1}{N_i}\sum_{j\in F_i}E(Y_j|\mathbb I_O)$, $V_i= \frac{1}{N_i}\sum_{j\in F_i}E(D_j|\mathbb I_T)$, $(u_i,v_i)$ are generated from the bivariate normal with mean 0's, variances 1's and correlation $\rho$; $X_1$ is generated as $N(1,1)$ and $X_2$ is generated as $N(0,1)$. We consider the case that $\alpha =1$. We generate a random social network of $n$
individuals as follows. Each individual $i$ has a degree independently drawn
from $N_{i}\in \{0,1,\cdots,10\}$ with equal likelihood on each degree. We
randomly choose $N_{i}$ of the other $n-1$ individuals as individual's
friends. The network is directed since $j$ can influence $i$ without
requiring, but not precluding, that $i$ influences $j$. We could 
simulate alternative network structures but the one adopted simplifies
the data generating process.

Recall that the BNEs are characterized by \Cref{bnes} and \Cref{bneo}. The parameters are set as $\beta_T=(-1,1,1)$, $\beta_O=(-1,1)$, $\alpha=1$, $\gamma=1$, $\delta=1$ and $\rho=(0.5,-0.5)$. We obtain the finite sample performance with sample sizes, $n=250,500, 1,000$ and 200 replications. Table C.1 and C.3 report the Average Biases and the Mean Squared Errors for $\rho=0.5$ and Table C.1 and C.3  report the Average Biases and the Mean Squared Errors for $\rho=-0.5$. Tables C.2 and C.4 report the performances of the estimators of the APTE.

\begin{table}[th]
\caption{Average Bias and Mean Squared Error}\label{double}
\centering
\begin{tabular}{r|cc|c|c|ccc|c|c}
\hline\hline
$n$ & \multicolumn{4}{c|}{Outcome} & \multicolumn{4}{c|}{Treatment}& \\ \hline
& \multicolumn{9}{c}{Average Bias} \\ \hline
& \multicolumn{2}{c|}{$\beta_{O}$} & $\gamma$ &$\delta$& \multicolumn{3}{c|}{$\beta^T$} & $\alpha$ &  $\rho$ \\ \hline
250 &-0.007 & 0.033 & 0.035& -0.021& -0.023&  0.019 & 0.040& -0.027 &-0.004\\ 
500 &0.012 & 0.016 & 0.004 &-0.034 &-0.010 & 0.025 & 0.003 &-0.001  &0.000\\ 
1,000&-0.008 & 0.003&  0.004&  0.013& -0.016& -0.004&  0.010& -0.007&  0.000\\\hline
%2,000& -0.001 & 0.002 & 0.003& -0.008 & 0.013 & 0.001 & 0.005& -0.009 & 0.002\\\hline
& \multicolumn{9}{c}{Mean Squared Error} \\ \hline
& \multicolumn{2}{c|}{$\beta_{O}$} & $\gamma$ &$\delta$& \multicolumn{3}{c|}{$\beta^T$} & $\alpha$ &  $\rho$ \\ \hline
250 &0.079 &0.023 &0.021& 0.174& 0.131 &0.026 &0.169 &0.194 &0.048\\ 
500 & 0.043& 0.010& 0.009 &0.094& 0.050& 0.011& 0.067& 0.069 &0.017\\ 
1,000&0.017& 0.005& 0.004 &0.039& 0.032& 0.006& 0.037 &0.048 &0.010\\
%2,000&0.006& 0.002& 0.002& 0.035& 0.005& 0.002& 0.019& 0.024& 0.005\\
\hline\hline
\end{tabular}%
\end{table}

\begin{table}[ht]
    \centering
        \caption{APTE}    \label{ame1}
    \begin{tabular}{c|ccc}
    \hline\hline
        n  & APTE true  &APTE Estimate &MSE \\\hline
        250 &0.746& 0.747& 0.002\\
        500&0.747 &0.743& 0.001\\
        1,000 &0.747& 0.746& 0.000\\
        %2,000&0.472& 0.475& 0.002\\
        \hline\hline
    \end{tabular}
\end{table}

\begin{table}[th]
\caption{Average Bias and Mean Squared Error}\label{double2}
\centering
\begin{tabular}{r|cc|c|c|ccc|c|c}
\hline\hline
$n$ & \multicolumn{4}{c|}{Outcome} & \multicolumn{4}{c|}{Treatment}& \\ \hline
& \multicolumn{9}{c}{Average Bias} \\ \hline
& \multicolumn{2}{c|}{$\beta_{O}$} & $\gamma$ &$\delta$& \multicolumn{3}{c|}{$\beta^T$} & $\alpha$ &  $\rho$ \\ \hline
250 &-0.014 & 0.028&  0.011 & 0.006&  0.019 & 0.022&  0.009 &-0.084 &-0.001\\ 
500 &-0.023 & 0.022 & 0.018 & 0.019& -0.021 & 0.021&  0.006 & 0.005 & 0.004\\ 
1,000&-0.014&  0.015&  0.011&  0.001& -0.022&  0.004 & 0.014&  0.004& -0.006\\\hline
%2,000& -0.001 & 0.002 & 0.003& -0.008 & 0.013 & 0.001 & 0.005& -0.009 & 0.002\\\hline
& \multicolumn{9}{c}{Mean Squared Error} \\ \hline
& \multicolumn{2}{c|}{$\beta_{O}$} & $\gamma$ &$\delta$& \multicolumn{3}{c|}{$\beta^T$} & $\alpha$ &  $\rho$ \\ \hline
250 &0.090& 0.024& 0.019& 0.175& 0.076 &0.040 &0.117 &0.206& 0.054\\ 
500 &  0.037& 0.010& 0.014 &0.079& 0.039 &0.019& 0.064 &0.082& 0.029\\ 
1,000&0.021& 0.006& 0.005& 0.042& 0.020& 0.008& 0.027 &0.038& 0.010\\
%2,000&0.006& 0.002& 0.002& 0.035& 0.005& 0.002& 0.019& 0.024& 0.005\\
\hline\hline
\end{tabular}%
\end{table}

\begin{table}[ht]
    \centering
        \caption{APTE}    \label{ame2}
    \begin{tabular}{c|ccc}
    \hline\hline
        n  & APTE true  &APTE Estimate &MSE \\\hline
        250 &0.542& 0.541& 0.001\\
        500&0.538& 0.537& 0.001\\
        1,000 &0.542 &0.541& 0.000\\
        %2,000&0.472& 0.475& 0.002\\
        \hline\hline
    \end{tabular}

\end{table}

The evidence suggests that the estimators work well in these settings noting that we only focus on the estimation of the model parameters and the APTE.

\end{doublespace}
\end{appendices}
\end{document}